\documentclass [aps, pre, amssymb, amsmath, showpacs, twocolumn]{revtex4}

\usepackage{graphicx}
\usepackage{dcolumn}
\usepackage{bm}

\begin{document}

\title{Classical Nonlinear Response of a Chaotic System:
Langevin Dynamics and Spectral Decomposition}
\author{Sergey V.  \surname{Malinin}$^a$}
\author {Vladimir Y. \surname{Chernyak}$^a$}
\email{chernyak@chem.wayne.edu}
\affiliation{$^a$Department of Chemistry, Wayne
State University, 5101 Cass Ave., Detroit, MI 48202\\}
\date{\today}

\begin{abstract}
We consider the classical response of a strongly chaotic Hamiltonian system. The spectrum of such a
system consists of discrete complex Ruelle-Pollicott (RP) resonances which manifest themselves in
the behavior of the correlation and response functions. We interpret the RP resonances as the
eigenstates and eigenvalues of the Fokker-Planck operator obtained by adding an infinitesimal noise
term to the first-order Liouville operator. We demonstrate how the deterministic expression for the
linear response is reproduced in the limit of vanishing noise. For the second-order response we
establish an equivalence of the spectral decomposition with infinitesimal noise and the long-time
asymptotic expansion for the deterministic case.

%\keywords{Hadamard-Gutzwiller model, classical response, chaos, spectroscopy, Liouville, curvature,
%Ruelle-Pollicott resonances, collective oscillations}
\end{abstract}


\maketitle

\section{Introduction}

Vibrational molecular motion is an interesting and important example of complex dynamics. Due to
nonlinearity, nonadiabatic effects and interactions with the environment, the behavior of such
systems is remarkably rich, and their theoretical description away from the equilibrium becomes a
challenge. In many cases, e.g. when the frequencies of the vibrational modes do not exceed the
temperature, a completely classical description is adequate. The dynamics far from equilibrium is
typically irregular and can exhibit chaotic features. In the extreme case of strong chaos all
trajectories in phase space are unstable, and the phase-space dynamics shows mixing features.

Spectroscopic methods represent powerful tools for obtaining detailed information on vibrational
dynamics. The outcome of spectroscopic experiments can be naturally interpreted in terms of optical
response functions, either in the time or in frequency domain. The multi-time (nonlinear) response
functions that constitute the main objects in multi-dimensional time-domain spectroscopy are
usually analyzed using their spectral decompositions, i.e. the Fourier transforms with respect to
the time intervals. These time intervals can be viewed as the time delays between the exciting
pulses, provided the latter are short compared to the typical time scales of the system dynamics.
Periodic features of response functions are routinely interpreted as signatures of periodic
motions. The central frequencies of diagonal and off-diagonal (cross) peaks are associated with
some coherent vibrations, whereas the peak shapes contain information on the system-bath coupling.
Smooth behavior of response functions in the case of almost harmonic vibration of the primary
system is achieved due to its coupling to a macroscopically large number of the bath motions. The
bath is often considered in the harmonic approximation: such model is known in the spectroscopic
literature as a multi-mode Brownian oscillator.

The situation is totally different in the case of chaotic motion of the primary system. The
spectrum of a strongly chaotic system is known to consist of so-called Ruelle-Pollicott (RP)
resonances that represent the eigenmodes of the Perron-Frobenius operator \cite{Ruelle86,Pollicott}.
The imaginary part of any particular resonance describes an
oscillating feature in the system correlations, whereas its real part is responsible for the
correlations decay. An individual RP resonance is not directly related to any particular periodic
motion, although their positions can be expressed in terms of all periodic orbits in a very
collective way via the dynamical $\zeta$-function. These collective resonances in chaotic systems
should be distinguished from the signatures of stable periodic motions.

Nonlinear response functions of quantum systems can be conveniently represented in terms of
spectral decompositions via the system's stationary states. These spectral decompositions that are
often referred to as Bloembergen's expressions allow to relate resonances in the spectroscopic
measurement data to the transitions between the stationary states. A natural question that arises
in the context of approaching the classical limit is what would be the classical counterpart of the
quantum Liouville space spectral decomposition for the linear and nonlinear response functions? In
quantum mechanics the system state can be described by a density matrix, whereas the evolution is
governed by the quantum Liouville operator. One can try to reach the classical limit by replacing
the quantum density matrix by the classical phase-space distribution using the Wigner
transformation. The quantum Liouville operator should be naturally replaced by its classical
counterpart.

This conceptually straightforward approach, however, faces certain major difficulties. Quantum
Liouville operators are second-order elliptic operators, whose eigenstates belong to certain
natural Hilbert spaces. In particular in the case of compact (restricted) coordinate spaces the
corresponding spectra are discrete, and finding spectral decompositions does not face conceptual
difficulties, at least on the physical level of rigor. The situation with the classical limit is
much more involved since the classical Liouville operator is a first-order non-elliptic operator
that describes the propagation of phase-space distributions along classical trajectories. In
particular any non-periodic classical trajectory generates a family of eigenstates of the classical
Liouville operator, concentrated on the trajectory with all possible eigenvalues. Such a spectrum
contains no useful information on the system relaxation.

A meaningful spectrum that provides detailed information on the relaxation of a strongly chaotic
system is known to be represented by the RP resonances. One of the ways to reproduce the RP
resonances as eigenvalues of the classical Liouville operator is based on the appropriate
simultaneous definition of the functional space where the operator acts. These functional Hilbert
spaces, often referred to as rigged spaces, are very different from ``standard'' Hilbert spaces
involved in spectral decompositions of quantum operators. They are usually chosen on the
case-to-case basis.

To avoid these difficulties we follow a more physical approach introduced in Ref.
\onlinecite{GNPT}. The approach is based on introducing weak Langevin noise, followed by
considering the limit of vanishing noise. It has several major advantages. First of all, this
describes a realistic situation since any system is at least weakly coupled to some environment,
and in many cases the system-bath coupling can be described on the level of Langevin noise. Second,
it allows to avoid dealing with a problem of choice of the appropriate Hilbert space. The
stochastic Langevin processes can be described by adding a diffusion operator to the Liouville
operator $\hat{L}$, which results in the Fokker-Planck operator $\hat{{\cal
L}}=-(\kappa/2)\nabla^{2}+\hat{L}$. The Fokker-Planck operator $\hat{{\cal L}}$ is a second-order
elliptic operator, and its spectral decomposition is free of the aforementioned difficulties that
arise in the case of classical Liouville operator. The RP resonances are obtained from the spectrum
of the Fokker-Planck operator by taking the limit $\kappa\to 0$. This allows to interpret
introducing infinitesimal noise as a regularization procedure similar to coarse graining.

The dependence of the eigenfunctions of the Fokker-Planck operator $\hat{{\cal L}}$ on the noise
strength $\kappa$ is nonanalytical. In the limit $\kappa\to 0$ of the vanishing noise strength the
smooth eigenfunctions turn into generalized functions (distributions). The rigged Hilbert spaces
are reproduced from the ``standard'' Hilbert spaces in the limit $\kappa\to 0$. They are spanned on
the generalized eigenfunction obtained from the eigenfunctions of $\hat{{\cal L}}(\kappa)$ by
applying the noiseless limit.

The calculation of the linear response function in terms of the generalized functions does not pose
a problem, since the expression involves an expansion of the smooth function over generalized
functions. The calculation of nonlinear response functions is essentially more complicated because
one needs to define expansions of the generalized functions over similar generalized functions. We
avoid dealing with the latter problem by performing all calculations for weak, yet finite noise,
followed by applying the limit $\kappa\to 0$ to the final expressions.

In our previous work \cite{MalininChernyak} we demonstrated the convergence of the second-order
response function for strongly chaotic systems. As an example of the strong chaos, we considered
free motion on a compact surface of constant negative curvature. The model has been studied since
more than a century and served as a prototype for quantum chaos
\cite{BalazsVoros,Gaspardbook,Gutzwiller}. The motion on a surface of genus $2$ is in particular
known as a Hadamard billiard. Using the dynamical symmetry (DS) of the system we found analytical
expressions for the response functions.  The long-time asymptotic expansion of the second-order
response function turned out to have the form of the double decomposition in the resonances that
appear in the linear response.

In the present work we find the decomposition of the linear and nonlinear response in
Ruelle-Pollicott resonances for this classical strongly chaotic model by introducing weak Langevin
noise. The spectral decomposition in this case is conceptually straightforward. We further
demonstrate that the decomposition coefficients converge in the noiseless limit $\kappa\to 0$, and
the resulting spectral decompositions reproduce the asymptotic expansions for the purely classical
response functions computed in our earlier work \cite{MalininChernyak}. To the best of our
knowledge, this is the first calculation of the classical nonlinear response for a chaotic flow
that uses spectral decomposition based on the noise regularization.

Our paper is organized as follows. In Sec. \ref{section:model} we summarize the geometry and
dynamical symmetry of free motion on a compact surface of constant negative curvature. Statistical
description of the response is presented in Section  \ref{section:noise}. We regularize the
dynamics by adding noise to the Liouville operator and derive general forms of the spectral
decompositions in Section \ref{section:decomposition-general}. The eigenstates of the resulting
Fokker-Planck operator are found in Section \ref{section:eigenstates}. Explicit forms of spectral
decompositions are obtained in Section \ref{section:linear} for the linear response and in Section
\ref{section:second-order} for the second-order response.

\section{Chaotic model system: Liouville-space picture and dynamical symmetry}
\label{section:model}

Following \cite{MalininChernyak} we consider free motion on a $2D$ compact surface $M^{2}$ of
constant negative curvature (Gaussian curvature). This strongly chaotic system is described by the
classical free-particle Hamiltonian
\begin{eqnarray}
\label{Hamiltonian}
H({\bm x},\zeta)=\frac{1}{2m} g^{ik}p_i p_k=\frac{\zeta^{2}}{2m}\,,
\end{eqnarray}
that depends on the absolute value $\zeta$ of the momentum $\bm p$  only. The curvature of the $2D$
configuration space is expressed in terms of the metric tensor $g^{ik}$. The Hamiltonian classical
dynamics preserves the smooth compact $3D$ manifold $M^{3}$ that corresponds to a fixed value of
energy. Points ${\bm x}\in M^{3}$ of the reduced phase space are described by  two coordinates $\bm
r \in M^{2}$ and the momentum direction angle $\theta$. Hereafter, we will use dimensionless units
so that the mass $m=1$ and the curvature $K=-1$. Despite the complexity of the flow due to its
chaotic nature, its strong dynamical symmetry (DS) enables an analytical treatment
\cite{MalininChernyak}.

The geodesic flow $\dot {\bm\eta}=\hat L{\bm\eta}$, where ${\bm\eta}=({\bm r},{\bm p})=({\bm
x},\zeta)$ denotes a point in phase space, is generated by the Liouville operator
\begin{eqnarray}
\label{Liouville-operator} \hat L\varsigma=\left\{H,\varsigma\right\}= \frac{\partial
H}{\partial\bm p}\frac{\partial\varsigma}{\partial\bm r}- \frac{\partial H}{\partial\bm
r}\frac{\partial\varsigma}{\partial\bm p} \,.
\end{eqnarray}
Hereafter, we identify any vector field with the corresponding first-order operator of the
derivative in the vector field direction. We denote by $\sigma_{1}$ the vector field that
determines the phase space velocity: $\hat L=\zeta\sigma_{1}$. We further introduce the second
vector field $\sigma_{z}=\partial/\partial\theta$ in the tangent space, and finally set
$\sigma_{2}=[\sigma_{1},\sigma_{z}]$  (see e.g. Refs. \onlinecite{Lang,MalininChernyak} for the
details). A simple local calculation yields $[\sigma_2,\sigma_z]=-\sigma_1$ and
$[\sigma_{1},\sigma_{2}]=-K\sigma_{z}$, which implies that in the constant negative curvature case
the vector fields $\sigma_{1}$, $\sigma_{2}$ and $\sigma_{z}$ form the Lie algebra $so(2,1)$. The
group $SO(2,1)$ action in the reduced phase space $M^{3}$ is obtained by integrating the $so(2,1)$
algebra action. DS with respect to the action of the group $G\cong SO(2,1)$ does not mean symmetry
in a usual sense, i.e. that the system dynamics commutes with the group action, but rather reflects
the fact that the vector field $\hat L$ that determines the classical dynamics is represented by an
element of the corresponding Lie algebra $so(2,1)$, whereas the stable and unstable directions of
our hyperbolic flow are determined by $\sigma_z\mp\sigma_2$. The algebra generators $\sigma_l$,
$l=1,2,z$ are anti-Hermitian, i.e. satisfy the equalities $\sigma^\dag_l=-\sigma_l$ with respect to
the natural scalar product $(\varphi,\psi)=\int d{\bm x}\varphi^*(\bm x;s)\psi(\bm x;s)$. In
addition, the Poisson bracket of two functions $f({\bm x},\zeta)$ and $g({\bm x},\zeta)$ reads
\begin{eqnarray}
\label{Poisson-bracket}
&&
\{f,g\}=
\\
\nonumber
&&
\frac{\partial f}{\partial\zeta}(\sigma_{1}g)-(\sigma_{1}f)\frac{\partial g}{\partial\zeta}
+\frac{1}{\zeta}\left((\sigma_{2}f)(\sigma_{z}g)-(\sigma_{z}f)(\sigma_{2}g)\right)\,.
\end{eqnarray}

The smooth action of $G=SO(2,1)$ in $M^{3}$ can be interpreted as that the space ${\cal H}$ of
smooth functions in $M^{3}$ constitutes a representation of $G$, which turns out to be a unitary
representation (see Refs. \onlinecite{Kirillov,Lang,Williams}), and therefore can be decomposed
into a direct sum of irreducible representations of $G$. Stated differently, any distribution in
the reduced phase space $M^{3}$ can be decomposed in irreducible representations. DS implies that
the distributions in different representations evolve independently. We focus on the principal
series representations of $SO(2,1)$ since only these provide experimentally interesting
contributions to the linear and second-order response \cite{MalininChernyak}. A principal series
representation ${\cal H}_s$ is labeled by an imaginary number $s$, $\mathrm{Im\,}s>0$.
Eigenfunctions $\psi_k(\bm x;s)$ of the momentum rotation operator $\sigma_z$, hereafter also
referred to as angular harmonics, form a convenient basis set in the irreducible representation
${\cal H}_s$. The functions $\psi_0(\bm x;s)$ do not depend on the momentum direction and can be
viewed as eigenfunctions of the Laplacian operator on our compact Riemann surface $M^{2}$
\cite{RobertsMuz,Williams,MalininChernyak}. The Laplacian eigenvalues provide a set of numbers
$s\in{\rm Spec}(M^{2})$ according to the equation $\nabla^2\psi_0(\bm x;s)=(s^2-1/4)\psi_0(\bm
x;s)$. The angular harmonics have the following properties:
\begin{align}
\label{sigma-act-irred0}
&
\sigma_{z}\psi_{k}({\bm x};s)=ik\psi_{k}({\bm x};s)\,,
\\
\nonumber & \sigma_{\pm}\psi_{k}({\bm x};s)=\left(\pm k+\frac{1}{2}-s\right)\psi_{k\pm 1}({\bm
x};s) \,,
\end{align}
where we introduced the raising and lowering operators $\sigma_{\pm}=\sigma_{1}\pm i\sigma_{2}$.
The operators $\sigma_{\pm}$ are anti-Hermitian conjugated: $\sigma_{+}^{\dagger}=-\sigma_{-}$.

The description of unitary representations of $G$  \cite{Kirillov,Lang} allows us to identify the
functions $\psi_{k}({\bm x};s)$ in reduced phase space with the corresponding functions on the
circle $\Psi_{k}(u)$ for any $s\in{\rm Spec}(M^{2})$. In the case of the principal series
$\mathrm{Re\,}s=0$, $\mathrm{Im\,}s>0$, and $\Psi_{k}(u)$ form an orthogonal normalized basis set
with the natural scalar product. Therefore, the normalized functions $\psi_{k}({\bm x};s)$ are
naturally associated with $\Psi_{k}(u)=\exp(iku)$. On the circle the algebra generators are
represented by (see \cite{MalininChernyak})
\begin{align}
\label{so(2,1)-circle}
&
\sigma_{z}=\frac{d}{du}, \quad \sigma_{1}=\sin u\frac{d}{du}+\frac{1-2s}{2}\cos u\,,
\\
&
\sigma_{2}=-\cos u\frac{d}{du}+\frac{1-2s}{2}\sin u\,.
\nonumber
\\
&
\sigma_{\pm}=\exp(\pm iu)\left(\mp i\frac{d}{du}+\frac{1-2s}{2}\right)\,.
\nonumber
\end{align}

Using the dynamical symmetry (DS) of the problem, one can calculate the response functions.
Two-point correlations that are related to the linear response via the fluctuation-dissipation
theorem have been also considered in Ref. \onlinecite{RobertsMuz}. Detailed calculations of the
response functions are presented in our previous paper \cite{MalininChernyak}. Since the zero
harmonics $\psi_0(\bm x;s)$ are momentum-independent, the expansion of the dipole $f(\bm r)$ in
irreducible representations has the form
\begin{eqnarray}
\label{f-expansion}
f=\sum_{s\in{\rm Spec}_{0}(M^{2})}B_{s}\psi_0(\bm x;s)
\end{eqnarray}
where only the principle series contributions are retained. Typically the dipole is represented by
a smooth distribution, and, therefore, only a small number $N_f$ of representations are relevant in
the expansion. For the sake of the presentation clarity we focus on the case $N_f=1$.
Generalizations to $N_f>1$ are straightforward and can be easily performed.

\section{Langevin Stochastic Dynamics: from Liouville to Fokker-Planck equation}
\label{section:noise}

\subsection{Liouville equation and response functions}
\label{subsection:response}

Strongly chaotic systems are characterized by complicated irregular dynamics \cite{chaosbook}.
Exponential divergence of initially close trajectories and mixing make the description in terms
of individual trajectories complicated and inadequate. An alternative picture,
based on Liouville representation of classical mechanics,
involves propagating phase space densities. Suppose that the system
initially occupies a certain region in the phase space which correspond to a range of initial phase
space variables. In the course of the evolution of a chaotic system, the shape of the region
becomes stretched along the unstable directions and contracted along the stable ones.

The response functions characterize the system response to a time-dependent external driving field
${\cal E}(t)$ and constitute the basic outcome of spectroscopic measurements. The external field is
coupled to the system via the polarization $f(\bm r)$, also often referred to as the dipole. The
function $f({\bm r})$ of the system coordinates is the classical counterpart of the polarization
operator. The total classical Hamiltonian of the system coupled to the driving field has the form
\begin{eqnarray}
H_T=H-f{\cal E}(t)\,.
\end{eqnarray}

The initial equilibrium phase space distribution $\rho_0$ starts to evolve once the external field
is turned on. The phase space density $\rho({\bm \eta},t)$ at time $t$ depends on the values of the
external field at preceding times. In most cases the observed signal at time $t$ is directly
related to the same polarization $f(\bm r)$, which describes the system-field coupling. Within the
classical mechanics formalism the observed polarization is given by $\int d\bm \eta\, f(\bm r)
\rho({\bm \eta},t)$. The evolution of the phase space distribution $\rho$ is governed by the
classical Liouville equation
\begin{eqnarray}
\label{Liouville-equation}
(\partial_t+\hat L)\rho={\cal E}(t)\{f,\rho\}\,,
\end{eqnarray}
where the Liouville operator $\hat L$ acts as the Poisson bracket with the Hamiltonian according to
Eq.~(\ref{Liouville-operator}).

The sign of the Poisson bracket is defined by a convention that $\{p_i,r_j\}=\delta_{ij}$ for
canonically conjugate coordinates $\bm r$ and momenta $\bm p$. Note that the Liouville operator is
a first-order linear differential operator. As stated earlier, it can be identified with the vector
field $\hat L=\zeta\sigma_1$ in the phase space. The Liouville operator is anti-Hermitian with
respect to the conventional scalar product, so that $\hat L^\dag =-\hat L$. This makes the
evolution operator $e^{-\hat L t}$ unitary and leads to the conservation of phase space volume
(Liouville theorem).

One can solve Eq.~(\ref{Liouville-equation}) iteratively considering the right-hand side as a small
perturbation and taking $\rho_0$ for the zero approximation. Then the phase space density, and
hence the response (observed signal) can be naturally represented in the form of the functional
Taylor expansion over the external field. The $n$-th order response function $S_n$ is defined as
the coefficient (more precisely, an integral kernel) in front of the $n$-th power of the driving
field. The response function $S_n$ depends on $n+1$ time moments, and if the unperturbed system is
initially at equilibrium, it actually depends on $n$ time intervals.

For the linear and second-order response functions we have the following expressions:
\begin{align}
\label{first-order}
&
S^{(1)}(t_1)
=\int d{\bm \eta}\,
f({\bm \eta})e^{-\hat L t_1}\{f({\bm \eta}),\rho_0\}\,,
\\
\label{second-order}
&
S^{(2)}(t_1,t_2)
=\int d{\bm \eta}\,
f({\bm \eta})e^{-\hat L t_2}
\{f({\bm \eta}),e^{-\hat L t_1}\{f({\bm \eta}),\rho_0\}\}\,.
\end{align}

Here $e^{-\hat Lt}$ is the the phase space density evolution operator, often referred to as the
Perron-Frobenius operator. We understand $e^{-\hat Lt} g({\bm \eta})$ as the solution $\rho({\bm
\eta},t)$ of the equation $(\partial_t+\hat L)\rho=0$ with the initial condition $\rho(\bm
x,0)=g(\bm x)$. Since $\hat L$ is a first-order differential operator, the solution can be readily
found using the method of characteristics. In other words, the first-order operator $\hat L$ allows
to write $e^{-\hat Lt} g({\bm \eta}) = g(e^{-\hat Lt}{\bm \eta})$.

\subsection{Langevin dynamics and Fokker-Plank equation}
\label{subsection:Fokker-Planck}

Phase space trajectories found from the Hamilton equations of motion are invariant with respect to
the time reversal. This property is easily observed in the behavior of integrable systems. Namely,
any integrable dynamics can be represented by a set of quasiperiodic motions, which implies
reversibility and recurrence in the values of physical observables. The behavior of chaotic systems
is quite different. Although described by the same formalism based on
Eqs.~(\ref{Liouville-operator}) and (\ref{Liouville-equation}) that possess time-reversal symmetry,
one observes obviously irreversible features such as relaxation phenomena.

The apparent paradox and its solution are well known in statistical mechanics. In the course of
evolution of a chaotic system, more and more fine features develop in the phase space distribution.
The distribution width decreases exponentially along the stable directions. At the same time,
physical quantities are represented by smooth functions in phase space. Therefore, fine features of
the distribution are actually irrelevant for the smooth observables. In particular, this results in
the exponential damping of the linear and nonlinear response functions in a strongly chaotic system
\cite{MalininChernyak}.

Only coarse grained properties of the phase space density remain relevant.
For instance, in the long-time asymptotic,
the strongly chaotic system may be found in the neighborhood of any point of the
phase space
with equal probability, which is reflected in the homogeneity of the stationary  phase space density.
A physically useful and meaningful definition of the evolution operator which relates initial and
final distributions must rely on some kind of regularization that eliminates unnecessary
distracting details. Coarse graining can be introduced either directly in phase space or,
alternatively, in the functional space of phase space distributions. Sometimes coarse graining is
inevitable, e.g. in computer simulations, where numerical errors determine the precision.

The finest scale of classical dynamics is limited by the onset of quantum effects. One of
well-known examples is the quasiclassical calculation of the entropy. Quantum effects were also
shown to remove unphysical power-law divergences in the nonlinear response functions of integrable
systems \cite{KryvohuzCaoPRL05,NL05}. It has been demonstrated that the limit $\hbar\to 0$ should
be taken after calculating the long-time asymptotics. This property resembles the characteristic
feature of our approach to the second-order response in a chaotic system, as presented below.

Besides, real physical systems are never isolated. The influence of an environment shows up as
noise or random forces at the level of equations of motion, and as unobservable degrees of freedom
and diffusive behavior of relevant variables in the statistical description. Although in many cases
interaction with the environment is sufficiently weak and can be neglected, it is often utilized as
a convenient and physically meaningful way to perform calculations.

The effect of the diffusion in the chaotic system is to regularize the long-time dynamics and
to introduce irreversible leveling of gradients in the stable direction. We regularize the operator
generating the density evolution by adding a small diffusive term in the form of the second-order
differential operator:
\begin{eqnarray}
\label{Fokker-Planck-operator}
\hat{\cal L} = \hat \sigma_1-\frac{\kappa}{2}\hat\sigma_z^2\,,
\end{eqnarray}
Since $\hat\sigma_z$ generates rotations of the momentum, $\kappa/2>0$ has the meaning
of the angular diffusion coefficient.
Factor $1/2$ is chosen for reasons of convenience. We will be interested in the case $\kappa\ll 1$.
As we will see, the limit $\kappa\to 0$ does exist in the spectral decomposition,
therefore in the limit the factor does not play any role.

The resulting Fokker-Planck operator $\hat{\cal L}$ defines the evolution of the phase space
density. This statistical description for the density is equivalent to the description in terms of
the Langevin equations for the dynamical variables. Such equations contain a random force which
tends to change the momentum direction keeping the energy constant.
This can be formally described by a stochastic Liouville operator
\begin{eqnarray}
\label{Liouville-stochastic}
\hat{L}_{st}(t)=\zeta\left(\hat{\sigma}_{1}+\gamma(t)\hat{\sigma}_{z}\right)
\end{eqnarray}
where $\gamma(t)$ is a random Gaussian Markovian process with the zero mean. The noise intensity is
determined by $\kappa$, and the two-point correlation function reads
\begin{eqnarray}
\label{gamma-correlation} \left\langle\gamma(t)\gamma(t')\right\rangle=\kappa\delta(t-t').
\end{eqnarray}

The addition of the diffusion term with the second derivative defines the spectrum of the resulting
operator in the space of smooth functions. The eigenfunctions become regular, differentiable
functions. As we will see, they are still singular at $\kappa=0$, and in the limit $\kappa\to 0$
they turn into generalized functions, which is natural for the case of a small parameter in front
of the highest derivative.

The mixing property of chaos makes the type of noise irrelevant. Random force exerted on any of the
phase space variables affects other variables if they are not fixed by conservation laws. Thus, in
the reduced phase space represented by the shell of constant energy the mixing leads to the fast
and irregular randomization of the position variables $\bm r$.

As a result, the Liouville equation (\ref{Liouville-equation}) that describes the phase space
density evolution in a system perturbed by an external field ${\cal E}$, is replaced by the
Fokker-Planck equation:
\begin{eqnarray}
\label{Fokker-Planck-equation}
(\partial_t+\zeta\hat{\cal L})\rho={\cal E}(t)\{f,\rho\}\,,
\end{eqnarray}
The evolution operator in the unperturbed system can be written symbolically as $e^{-\zeta\hat{\cal
L}t}$, in a full analogy with the noiseless case. Similarly, the iterative solution of
Eq.~(\ref{Fokker-Planck-equation}) yields the phase space density in a form of an expansion in
powers of ${\cal E}$. The response functions are obtained from Eqs.~(\ref{first-order}) by
replacing $\hat L$ with $\zeta\hat{\cal L}$:
\begin{align}
\label{first-order-general1}
&
S^{(1)}(t)
=\int d\zeta\,\zeta \langle fe^{-{\zeta\cal L} t}
f_- \rho_0\rangle\,,
\\
\label{second-order-general1}
&
S^{(2)}(t_1,t_2)
=\int d\zeta\,\zeta \langle fe^{-{\zeta\cal L} t_2}
f_- e^{-{\zeta\cal L} t_1}f_- \rho_0\rangle\,,
\end{align}
where angular brackets stand for the integral over the reduced phase space,
and we defined the action of the operator
$\hat f_-$ on a function $g(\bm\eta)$ as the Poisson bracket
of $f$ and $g(\bm\eta)$ so that $\hat f_-g=\{f,g\}$.

\section{Spectral decomposition of response functions: General formalism}
\label{section:decomposition-general}

In the well-known situation with quantum response, the response functions can be readily
represented in the form of spectral decompositions, since the infinitesimal evolution is determined
by a Hermitian operator. The conjugation property with respect to a simple scalar product
facilitates expansions in the basis of the eigenstates.

We have a natural scalar product for functions in $M^3$ which allowed to implement the unitary
presentations in $3D$ space by functions on the circle. With respect to this scalar product
$\hat{L}$ is an anti-Hermitian first-order differential operator. However, the Fokker-Planck
operator $\hat{\cal L}$ is obtained by adding a second-order Hermitian contribution to $\hat{L}$.
This makes the resulting operator $\hat{\cal L}$ neither Hermitian nor anti-Hermitian. Namely, its
adjoint is
\begin{eqnarray}
\label{adjoint}
\hat{\cal L}^\dag=-\sigma_1-\frac{\kappa}{2}\sigma_z^2\,.
\end{eqnarray}
It seems that there is no natural choice of the scalar product that would make our Fokker-Planck
operator simply related to its adjoint.

The expressions for the response functions include integrations over the reduced phase space. The
convolution $\langle \varphi(\bm x) \psi(\bm x)\rangle\equiv\int d\bm x\,\varphi(\bm x)\psi(\bm x)$
of two functions $\varphi(\bm x)$ and $\psi(\bm x)$ can be recast in the form of the natural scalar
product as $\langle \varphi \psi\rangle = (\varphi^*, \psi)$.

As shown by the construction of eigenmodes in Section \ref{section:eigenstates}, the operator
$\hat{\cal L}$ is diagonalizable in ${\cal H}_s$, and the Jordan-block structures that are possible
in a general case do not appear in our spectral decompositions. Therefore, any function in ${\cal
H}_s$ can be expanded in the eigenfunctions $\varphi_\lambda(\bm x;s)$ that obey the equation
\begin{eqnarray}
\label{Lvarphi}
\hat{\cal L}\varphi_\lambda=\lambda\varphi_\lambda\,.
\end{eqnarray}
The expansion coefficients are linear functionals of $\varphi(\bm x;s)$ and hence can be
represented by scalar products of $\varphi(\bm x;s)$ with certain functions denoted by
$\tilde\varphi_\lambda(\bm x;s)$:
\begin{eqnarray}
\label{varphi-expansion}
\varphi(\bm x;s)=\sum_\mu
\langle \tilde\varphi^*_\mu(\bm x;s) \varphi(\bm x;s)\rangle \varphi_\mu(\bm x;s)\,.
\end{eqnarray}
It follows from the expansion of $\varphi_\lambda(\bm x;s)$ and $\hat{\cal L}\varphi_\lambda(\bm
x;s)$ that the functions $\tilde\varphi_\lambda(\bm x;s)$ are the eigenfunctions of the operator
$\hat{\cal L}^\dag$ which satisfy the following properties:
\begin{align}
\label{tildevarphi-properties}
 &
 \hat{\cal L}^\dag \tilde\varphi_\lambda(\bm x;s)=\lambda^* \tilde\varphi_\lambda(\bm x;s)\,,
 \\
 &
 \langle \tilde\varphi^*_\mu(\bm x;s) \varphi_\lambda(\bm x;s) \rangle =\delta_{\mu\lambda}\,.
\end{align}

We focus our detailed treatment on the simplest case $N_f=1$ of a single irreducible representation
contributing to the dipole moment $f(\bm r)$. This corresponds to a single contribution in the
expansion (\ref{f-expansion}), and below we imply $f=\psi_0(\bm x,s)$. A generalization to an
arbitrary linear combination of $N_f>1$ such terms with different $s$ is straightforward for both
response functions under consideration \cite{MalininChernyak}.

We can successively apply the expansion procedure (\ref{varphi-expansion})
and obtain any response or correlation function as a spectral decomposition.
Based on Eq.~(\ref{first-order-general1})
we get the following form for the linear response function:
\begin{align}
\label{first-order-general2}
S^{(1)}(t)
=
&
\int d\zeta\,\zeta \sum\limits_{\lambda}e^{-\lambda\zeta t}
\langle f\varphi_\lambda\rangle
\langle \tilde\varphi_\lambda f_- \rho_0\rangle\,.
\end{align}

For the calculation of the second-order response starting from Eq.~(\ref{second-order-general1}),
we in addition introduce a double expansion in the basis of the angular harmonics that are defined
by Eqs.~(\ref{sigma-act-irred0}):
\begin{align}
\label{second-order-general2}
&
S^{(2)}(t_1,t_2)
=
\int d\zeta\,\zeta \sum\limits_{mn}\sum\limits_{\mu\lambda}e^{-\mu\zeta t_2}
\times
\\
\nonumber
&
\langle f\varphi_\mu\rangle
\langle \tilde\varphi_\mu \psi_n \rangle
\langle \psi^*_n f_- \psi_m \rangle \langle\psi^*_me^{-\lambda\zeta t_1}\varphi_\lambda\rangle
\langle \tilde\varphi_\lambda f_- \rho_0\rangle\,.
\end{align}

The double expansion in the angular harmonics yields a geometric matrix element
$\langle \psi^*_n \psi_0 \psi_m \rangle$ studied in our previous work \cite{MalininChernyak}.
Note that $f_-$ in the middle angular brackets contains a derivative $\partial/\partial\zeta$
acting on all functions of the momentum to the right of it.
This complication occurs due to the absence of the integration over $\zeta$
in the angular brackets which correspond to projections onto the basis set in ${\cal H}_s$.

\section{Eigenstates of Fokker-Plank operator}
\label{section:eigenstates}

As stated earlier the Fokker-Planck operator (\ref{Fokker-Planck-operator}) is neither Hermitian
nor anti-Hermitian with respect to any natural scalar product and, therefore, not necessarily
diagonalizable. In what follows, we show that it actually is: We apply the representation on the
circle to find its eigenstates and demonstrate that they constitute a basis set.

The eigenfunctions  and eigenvalues of $\hat{\cal L}$ can be found by solving the $1D$ eigenvalue
problem
\begin{eqnarray}
\label{Eq-general}
\hat{\cal L} \Phi_\lambda =\lambda \Phi_\lambda\,.
\end{eqnarray}
on a circle with the Fokker-Planck operator [see Eqs. (\ref{so(2,1)-circle}) and
(\ref{Fokker-Planck-operator})]
\begin{eqnarray}
\label{operator1}
 \hat{\cal L} = -\frac{\kappa}{2}\frac{d^2}{du^2}+\sin u \frac{d}{du}
 +\frac{1-2s}{2}\cos u\,.
\end{eqnarray}

The first derivative in the operator can be eliminated by redefining the functions
\begin{eqnarray}
\label{Phixirelation}
\Phi_\lambda(u)=e^{-\frac{\cos u}{\kappa}}\xi_\lambda(u)\,.
\end{eqnarray}
In terms of the functions $\xi_\lambda(u)$, the eigenvalue problem (\ref{Eq-general}) assumes the
form of a stationary Schr\"odinger equation
\begin{eqnarray}
\label{Schroedinger}
\hat {\cal H} \xi_\lambda(u) = \lambda \xi_\lambda(u)
\end{eqnarray}
with the effective Hamiltonian
\begin{align}
\label{Hamiltonian-on-circle}
%\\
\hat {\cal H}=-\frac{\kappa}{2}\frac{\partial^2}{\partial u^2}+\frac{\sin^2 u}{2\kappa}
-s\cos u \,.
\end{align}

The Hamiltonian describes a quantum particle in a complex-valued potential on the circle. The
operator in not Hermitian because of the imaginary part of the potential. However, since the
Hamiltonian does not contain first derivatives, we can define a symmetric (non-Hermitian) scalar
product $V(\xi,\phi)=\int du\, \xi(u) \phi(u)$ so that the Hamiltonian is self-adjoint with respect
to it: $V(\hat {\cal H} \xi,\phi)=V(\xi,\hat {\cal H}\phi)$. Eigenfunctions of $\hat {\cal H}$
corresponding to different eigenvalues are orthogonal and can be normalized:
\begin{eqnarray}
\label{scalar-product-real}
\int\limits_0^{2\pi} du\,\xi_\lambda(u)\xi_\mu(u)=\delta_{\lambda\mu}\,.
\end{eqnarray}

The Hamiltonian $\hat {\cal H}$ also possesses certain symmetries that simplify the analysis.
First, it is invariant with respect to the sign change of $u$, and therefore all its eigenfunctions
are either even or odd functions of $u$. As we will see, only even eigenstates contribute to the
spectral decompositions of response functions. Therefore, we present the detailed expressions only
for even eigenstates on the half-circle $0<u<\pi$.

Since the potential contains a nonzero imaginary part, its eigenvalues can be complex. The second
symmetry involves the complex conjugation of the Hamiltonian: ${\cal H}^*(u)={\cal H}(u+\pi)$.
Consequently, the eigenfunctions corresponding to the complex conjugate eigenvalues are related by
\begin{align}
\label{conjugation-symmetry}
\xi_{\lambda^*}(u)=\left(\xi_\lambda(u+\pi)\right)^*\,.
\end{align}

Detailed analysis of the Schr\"odinger equation in the weak noise case $\kappa \ll 1$ is presented
in Appendix \ref{appendix:Eigenmodes}. The small parameter $\kappa$ allows to solve the
Schr\"odinger equation using the WKB method. The imaginary part of the potential energy is small
compared to its real part. This supports the use of such terms as the ``minimum'' of the potential
and the ``under the barrier'' wave function. Since the potential minima become deeper for smaller
$\kappa$, the eigenfunctions are concentrated near $u=0$ or $u=\pi$. These states do not mix
together since $s$ is imaginary.

Due to the compact nature of the circle, the spectrum of the Fokker-Planck operator is discrete.
Its real part is positive and unbounded from above. For such large energies that
$\mathrm{Re}\,\lambda\gg \kappa^{-1}$, the spectrum can be adequately approximated by
$\lambda_\nu\sim \kappa\nu^2/2$ (with the eigenfunctions $\xi_{\lambda_{\nu}}\sim e^{i\nu u}$) of a
free particle on the circle.

In the limit $\kappa \to 0$ even low-energy eigenstates of the first set are concentrated near
$u=0$ and have energies
\begin{align}
\label{eigenvalues}
\lambda_\nu=\nu-s+\frac{1}{2}\,,
\end{align}
with nonnegative even numbers $\nu$ ($\nu=0,2,4,\ldots$). This discrete equidistant spectrum  of
the Fokker-Planck operator in the noiseless limit $\kappa\to 0$ is a quite fascinating property.
Infinitesimal noise regularizes the Liouvillian dynamics to yield a physical spectrum in the space
of smooth functions.

The normalized eigenfunctions that correspond to the specified above eigenvalues are concentrated
near $u=0$:
\begin{align}
\label{harmonic-zero}
&
\xi_{\lambda_\nu}(u)=A_{0,\nu}e^{-u^2/2\kappa}H_\nu\left(\frac{u}{\sqrt{\kappa}}\right)\,,
\\
\label{A0}
&
A_{0,\nu}=(\pi\kappa)^{-1/4}(2^\nu \nu!)^{-1/2}\,,
\end{align}
where $H_\nu$ are Hermite polynomials. The eigenfunctions $\Phi_{\lambda_\nu}(u)=e^{\cos
u/\kappa}\xi_{\lambda_\nu}(u)$ are singular at $u=0$ in the limit $\kappa\to 0$, as they should
form a basis set of distributions with large gradients in the stable direction.

In the limit $\kappa\to 0$ the eigenvalues in Eq. (\ref{eigenvalues}) are determined by the single
potential minimum at $u=0$. The form (\ref{harmonic-zero}) of the eigenfunctions is appropriate in
the region where the potential may be approximated by a harmonic well. Outside the region, at
$|u|\gtrsim 1$, the values of the eigenfunction are exponentially small and do not influence its
normalization.

A more rigorous treatment in Appendix provides the WKB approximation for the even eigenfunctions
$\xi_\nu(u)$ on the whole circle. Under the potential barrier, if $u\gtrsim\kappa^{1/2}$ and
$(\pi-u)\gtrsim\kappa^{1/2}$ the approximate solution at the energy $\lambda_\nu=\nu-s+1/2$ is
found to be
\begin{align}
\label{under-barrier-positive-half} & \xi(u)=A_{2,\nu} 2^{s-\frac{1}{2}} \left(\sin
\frac{u}{2}\right)^{\nu}\! \left(\cos\frac{u}{2}\right)^{s-\nu-\frac{1}{2}} e^{\frac{\cos
u}{\kappa}}\,,
\\
\label{A2}
&
A_{2,\nu}=A_{0,\nu} e^{-\frac{1}{\kappa}} 2^{\nu+\lambda} \kappa^{-\frac{\nu}{2}}\,.
\end{align}
In the vicinity of $u=\pi$, at $\pi-u\ll 1$, we obtain:
\begin{eqnarray}
\label{harmonic-pi}
\xi(u)
&=&
A_1e^{-(u-\pi)^2/2\kappa}H_{\nu_1}\left(\frac{u-\pi}{\sqrt{\kappa}}\right)
\\
\nonumber
&
+&
B_1e^{(u-\pi)^2/2\kappa}H_{-\nu_1-1}\left(i\frac{u-\pi}{\sqrt{\kappa}}\right)\,,
\end{eqnarray}
where
\begin{align}
\label{nu1-0}
&
\nu_1=\nu-2s=\lambda-\frac{1}{2}-s\,,
\\
& A_1=-B_1 i 2^{-\nu_1} \pi^{-1/2}  \Gamma(-\nu_1)\cos\frac{\pi\nu_1}{2}\,,
\\
&
B_1=A_{2,\nu} e^{-\frac{1}{\kappa}} 2^{\nu_1+\lambda+1} \kappa^{-\frac{\nu_1+1}{2}}
e^{\frac{i\pi(\nu_1+1)}{2}}\,.
\end{align}

Even low-lying eigenstates of the second set concentrated near $u=\pi$ have eigenenergies
$\lambda^*_\nu=\nu+s+1/2$, and their eigenfunctions can be obtained by using
Eq.~(\ref{conjugation-symmetry}) which follows from the symmetry of the Hamiltonian.

Eqs.~(\ref{harmonic-zero})-(\ref{nu1-0}) provide a zero-order approximation for the eigenfunction
of the Hamiltonian (\ref{Hamiltonian-on-circle}) on the circle. In the limit $\kappa\ll 1$ one can
distinguish two types of corrections to the specified above eigenvalues and eigenfunctions of
low-lying states. The first one is due to the omitted terms in the potential in the vicinity of its
minima. Such corrections are important in the treatment of states with $\nu>0$. Corrections of the
second type originate from the tunneling through the potential barriers and therefore are
exponentially small $\sim e^{-2/\kappa}$.

Details of computing the eigenfunctions are presented in Appendix \ref{appendix:Eigenmodes}.
It turns out that a straightforward calculation
of the higher-mode contributions to the spectral decomposition requires knowledge of eigenfunctions
with increasing accuracy in $\kappa$. In the next two sections we present a more elegant approach
for which the approximation given by Eqs.~(\ref{harmonic-zero})-(\ref{nu1-0}) is sufficient.

\section{Spectral decomposition for linear response}
\label{section:linear}

Spectral decompositions of the response functions in the eigenmodes of the evolution operator are
obtained in Eqs. (\ref{first-order-general2}, \ref{second-order-general2}) in the general form. The
expressions are not really useful for the calculation, since it would involve several
three-dimensional integrations that involve the functions not explicitly known. Nevertheless, the
calculation can be carried out by implementing the representation in terms of functions on the
circle.

We start with the spectral decomposition of the linear response, and employ the correspondence
between the functions $\varphi$ in $M^3$ and $\Phi(u)$ on the circle, to obtain the following
expansion:
\begin{eqnarray}
\label{tildePhi} \Phi(u)=\sum_\lambda \langle \tilde\varphi^*_\lambda \varphi \rangle
\Phi_\lambda(u)\,.
\end{eqnarray}
Using the representation $\Phi_\lambda(u)=e^{-\cos u/\kappa}\xi_\lambda(u)$ and the orthonormality
of the functions $\xi_\lambda(u)$ with respect to the symmetric scalar product
(\ref{scalar-product-real}), we can specify a rule to find the projection of $\varphi$ on
$\varphi_\lambda$:
\begin{eqnarray}
\label{scalarproductsequ}
\langle \tilde\varphi^*_\lambda \varphi \rangle
=\int\limits_{-\pi}^\pi du\,  \Phi(u) e^{\frac{\cos u}{\kappa}}\xi_\lambda(u)\,.
\end{eqnarray}
This implies that $\tilde\varphi_\lambda$ is implemented on the circle by the function
\begin{eqnarray}
\label{tildePhixirelation}
\tilde\Phi_\lambda(u)=2\pi e^{\frac{\cos u}{\kappa}}\xi^*_\lambda(u)\,.
\end{eqnarray}

To proceed with the calculation of the response functions, we introduce a notation
\begin{eqnarray}
\label{R-definition}
R_{\lambda,n}\equiv\langle \tilde\varphi^*_\lambda \psi_n \rangle=
\int\limits_{-\pi}^\pi du\, e^{inu}e^{\frac{\cos u}{\kappa}}\xi_\lambda(u)
\end{eqnarray}
for the coefficient of the expansion of $\psi_n(\bm x;s)$ in $\varphi_\lambda(\bm x;s)$.
The expansion of $\varphi_\lambda(\bm x;s)$ in angular harmonics $\psi_n(\bm x;s)$ contains
coefficients
\begin{eqnarray}
\label{R-conjugate-integral}
\langle \psi^*_n  \varphi_\lambda \rangle =
\int\limits_{-\pi}^\pi \frac{du}{2\pi}\, e^{-inu}e^{-\frac{\cos u}{\kappa}}\xi_\lambda(u)\,.
\end{eqnarray}
The symmetry property (\ref{conjugation-symmetry}) allows to relate them to $R_{\lambda,n}$:
\begin{eqnarray}
\label{R-conjugate}
\langle \psi^*_n  \varphi_\lambda \rangle =
\frac{(-1)^n}{2\pi}(R_{\lambda^*,n})^*\,.
\end{eqnarray}

Therefore, the convolution in the first angular brackets in Eq.~(\ref{first-order-general2}) is
equal to $(R_{\lambda^*,0})^*/2\pi$. The action of the $f_{-}$ in the last angular brackets,
represented by a Poisson bracket (\ref{Poisson-bracket}), creates the following convolution
\begin{eqnarray}
\langle \tilde\varphi^*_\lambda \sigma_1 \psi_0 \rangle =
\frac{1-2s}{2}\int\limits_{-\pi}^\pi \frac{du}{2\pi}\, \cos u\,
e^{\frac{\cos u}{\kappa}}\xi_\lambda(u)\,,
\end{eqnarray}
where we have used the representation of $\sigma_1$ on the circle given by
Eq.~(\ref{so(2,1)-circle}). Bearing in mind that $\xi_\lambda$ is the eigenfunction of Hamiltonian
(\ref{Hamiltonian-on-circle}) and integrating by parts, we find the following relation:
\begin{eqnarray}
\langle \tilde\varphi^*_\lambda \sigma_1 \psi_0 \rangle =
\lambda R_{\lambda,0}\,.
\end{eqnarray}

We conclude that the linear response function can be recast in the form of the time derivative of
the two-point correlation function, in agreement with the fluctuation-dissipation theorem (FDT):
\begin{align}
\label{linear-response1}
S^{(1)}(t)=
\frac{\partial}{\partial t}
\sum\limits_\lambda\int\limits_0^\infty d\zeta\,
{\cal Q}_{\lambda,0} e^{-\zeta \lambda t}\frac{\partial \rho_0}{\partial \zeta}\,,
\end{align}
where we introduced
\begin{align}
\label{cal-Qn}
{\cal Q}_{\lambda,n}=\frac{(-1)^n}{2\pi} R_{\lambda,0} (R_{\lambda^*,n})^*\,.
\end{align}

Following the approach developed in Ref. \onlinecite{MalininChernyak} for the purely deterministic
situation, we introduce the matrix elements of the evolution operator between $n$-th and zero
harmonics:
\begin{align}
\label{cal-A-definition}
{\cal A}_n(\zeta t;s)=\int d\bm x\, \psi^*_n(\bm x;s)e^{-\hat {\cal L} t}\psi_0(\bm x;s)
\,.
\end{align}
Spectral decomposition
\begin{align}
\label{cal-An-series}
&
{\cal A}_n(\zeta t;s)=\sum\limits_\lambda {\cal Q}_{\lambda,n} e^{-\zeta \lambda t}
\end{align}
appears in the response functions of the system with noise.

As we show below, the noiseless limit $\kappa\to 0$ of the series ${\cal A}_n(t;s)$ reproduces the
matrix element $A_n(t;s)$ of the deterministic evolution operator, calculated in Ref.
\onlinecite{MalininChernyak}. These series and their coefficients ${\cal Q}_{\lambda,n}$ play an
important role in the calculation of the second-order response function.

We are now in a position to proceed with an explicit calculation of the coefficients in the
spectral decomposition of the linear response function. We focus on the first set of modes with the
energies $\lambda_\nu=\nu-s+1/2$, whose eigenfunctions are concentrated in the neighborhood of
$u=0$ for $\kappa \ll 1$. The coefficients ${\cal Q}_{\lambda,0}$ for the other set of modes, with
the energies $\lambda^*_\nu=\nu+s+1/2$, can be then easily found by employing the symmetry,
described by Eq.~(\ref{conjugation-symmetry}):
\begin{align}
\nonumber
&
{\cal Q}_{\lambda^*,0}= ({\cal Q}_{\lambda,0})^*\,,
\end{align}
which also reflects the fact that the response function is real.

The first two terms in the spectral expansion originate from the lowest RP resonances with the
energies $\lambda_0=1/2-s$ and $\lambda^*_0=1/2+s$. The eigenfunction with $\lambda_0=1/2-s$,
concentrated in the vicinity of $u=0$, is given by
$\xi_{\lambda_0}(u)=(\pi\kappa)^{-1/4}\exp(-u^2/2\kappa)$. The first integral, $R_{\lambda_0,0}$,
is calculated without much effort. This is done by noticing that the main contribution to the
integral comes from $|u|\lesssim \sqrt{\kappa}$. Expanding $\cos u= 1-u^2/2$ in the exponential
results in a Gaussian integral:
\begin{align}
R_{\lambda_0,0}=\int\limits_{-\pi}^\pi du\, e^{\frac{\cos u}{\kappa}}\xi_{\lambda_0}(u)=
e^{\frac{1}{\kappa}} A^{-1}_{0,0}\,,
\end{align}
where the normalization factor $A_{0,0}$ is related to $\kappa$ via Eq.~(\ref{A0}).

The situation, however, is more complicated for higher modes with $\nu>0$. Due to the orthogonality
of the eigenfunctions $\xi_{\lambda_0}$ and $\xi_{\lambda_\nu}$ with $\nu>0$, the integral vanishes
if we employ the lowest-order approximation for the wavefunctions:
\begin{align}
%\label{zero-mode-total-integral}
\int\limits_{-\pi}^{\pi}du\,e^{\frac{\cos u}{\kappa}}\xi_\lambda(u)\approx
%\\
A_{0,\nu}e^{\frac{1}{\kappa}}
\!\!\!
\int\limits_{|u|\lesssim \sqrt{\kappa}}du\,e^{-\frac{u^2}{\kappa}}
H_\nu\left(\frac{u}{\sqrt{\kappa}}\right)=0\,.
\end{align}
This actually means that more accurate expressions are needed to find the first nonvanishing term.
Note that one should use better approximations for both the exponential $e^{\cos u/\kappa}$ and the
wave function. The corrections to the latter are computed by applying the standard quantum
mechanical perturbation theory to the Schr\"odinger equation. The procedure is feasible for few low
modes only, since the calculation for the higher modes requires higher orders of the perturbation
theory.

However, it is possible to overcome this difficulty and find an explicit expression for
$R_{\lambda_\nu,0}$ for all $\nu$ by using the following trick. We notice that the dominant
contribution to the integral
\begin{align}
\label{integral1-final}
&
\int\limits_{-\pi}^{\pi}du\,e^{\frac{\cos u}{\kappa}}(\cos u-1)^{\nu/2}\xi_{\lambda_\nu}(u)=
2^{-2\nu}(-1)^{\nu/2}A_{0,\nu}^{-1}e^{1/\kappa}
\,.
\end{align}
does not involve higher-order perturbative calculations described above. The integral is dominated
by the region $|u|\lesssim\sqrt{\kappa}$, where we can approximate $(\cos
u-1)^{\nu/2}=(-1/2)^{\nu/2}u^\nu$ and expand it in the Hermite polynomials $H_j(u/\sqrt{\kappa})$
with $j=0,2,\ldots,\nu$. All terms except for the last one vanish due to orthogonality of the
Hermite polynomials. Further approximations lead to negligible corrections in the limit $\kappa\to
0$. We proceed by introducing
\begin{align}
\label{P-definition}
&
P_{j,\lambda}=\int\limits_{-\pi}^{\pi}du\,e^{\frac{\cos u}{\kappa}}(\cos u-1)^j
\xi_{\lambda}(u)
\,.
\end{align}
Our goal is to calculate the integral $R_{\lambda_\nu,0}=P_{0,\lambda_\nu}$. This will be achieved
by expressing it via the known integral $P_{\nu/2,\lambda_\nu}$ for $\kappa\ll 1$. Consider
$\lambda P_{j,\lambda}$. Making use of $\lambda \xi_{\lambda}=\hat{\cal H}\xi_{\lambda}$, we
integrate in Eq.~(\ref{P-definition}) by parts, and neglect higher-order terms in $\kappa$. This
results in the recurrence relation
\begin{align}
\label{P-recurrence}
&
P_{j,\lambda}=\frac{j-s+\frac{1}{2}}{\lambda-2j+s-\frac{1}{2}}P_{j+1,\lambda}\,,
\end{align}
which being applied $\nu/2$ times and followed by setting $\lambda=\lambda_\nu$ yields:
\begin{align}
&
P_{0,\lambda_\nu}=\frac{2^{-\nu/2}\Gamma\left(\frac{\nu+1-2s}{2}\right)}
{(\nu/2)!\Gamma\left(\frac{1-2s}{2}\right)}P_{\nu/2,\lambda_\nu}\,,
\end{align}
and we arrive at
\begin{align}
\label{R-result}
&
R_{\lambda_\nu,0}=
e^{\frac{1}{\kappa}}\kappa^{\nu/2} A_{0,\nu}^{-1}2^{-2\nu}
\frac{(-1)^{\nu/2}\Gamma\left(\frac{\nu+1-2s}{2}\right)}{(\nu/2)!\Gamma\left(\frac{1-2s}{2}\right)}
\,.
\end{align}

The second integral $\langle\psi_0\varphi_{\lambda_\nu} \rangle$ in the linear response is
expressed via $(R_{\lambda_\nu^*,0})^*$ [see Eqs. (\ref{R-conjugate-integral}) and
(\ref{R-conjugate})]. Making use of the complex conjugation symmetry specified in
Eq.~(\ref{conjugation-symmetry}) this can be recast as
\begin{align}
\label{R0-conjugate-integral}
(R_{\lambda_\nu^*,0})^*=\int\limits_{-\pi}^\pi du\, e^{-\frac{\cos u}{\kappa}}\xi_{\lambda_\nu}(u)
\,.
\end{align}
The obtained integral looks more complicated compared to $R_{\lambda_\nu,0}$, even for the lowest
resonance with $\nu=0$. We first point out the problem and present its straightforward solution for
$\nu=0$. Afterwards we apply a more elegant method, similar to the one developed above, which
provides the solution for all modes.

Before starting we emphasize that, due to the eigenfunction symmetry, the discussion can be limited
to the interval $0<u<\pi$. Three different analytical approximate forms for the eigenfunction near
$u=0$, $u=\pi$, and under the barrier are specified in Eqs. (\ref{harmonic-zero}) - (\ref{nu1-0}).
As opposed to the case of $R_{\lambda_\nu,0}$, the integrand in $(R_{\lambda_\nu^*,0})^*$ does not
experience exponential dependence on $u$, and the exponential part of its dependence on $\kappa$ in
the contributions from all three regions amounts to a common factor $\exp(-1/\kappa)$.

In the region $\sqrt{\kappa}\ll (\pi-u)\ll 1$, where the approximate forms, given by
Eqs.~(\ref{under-barrier-positive-half}) and (\ref{harmonic-pi}) match, the solution behaves $\sim
(\pi-u)^{-\nu-1+2s}$. This region, where the two solutions match, provides an essential
contribution to the integral. We can break up the integral into two parts by introducing an
intermediate point $\pi-u_1$ that belongs to the matching region. Since for small $\kappa \ll 1$
one can always choose $\kappa^{1/2}\ll u_1 \ll 1$, the approximate forms provide good
approximations to the left and to the right of the intermediate point, respectively. We intend to
show that the sum of two integrals is independent of $u_{1}$.

The contribution from the region nearby $u=\pi$ has the form:
\begin{align}
\label{other-minimum-integral}
&
\int\limits_{\pi-u_1}^\pi du\, e^{-\frac{\cos u}{\kappa}}\xi_\lambda(u)=
\\
\nonumber
&
e^{\frac{1}{\kappa}} \sqrt{\kappa} \int\limits_{0}^{u_1/\sqrt{\kappa}} dz\,
(A_1 e^{-z^2}H_{\nu_1}\left(z\right) + B_1 H_{-\nu_1-1}\left(iz\right))  =
\\
\nonumber
&
-B_1 e^{\frac{1}{\kappa}} \sqrt{\kappa} e^{-i\pi\nu_1/2}
\frac{H_{-\nu_1}\left(iu_1/\sqrt{\kappa}\right)-H_{-\nu_1}\left(-iu_1/\sqrt{\kappa}\right)}
{4\nu_1 \sin(\pi\nu_1/2)
%\left(e^{-i\pi\nu_1/2}-e^{i\pi\nu_1/2}\right)
}\,.
\end{align}
We chose $u_1$ so that $\kappa^{1/2}\ll u_1\ll \kappa^{1/4}\ll 1$ and retained first two terms in
the expansion of $\cos u$. The integral is calculated by expressing $e^{-z^2}H_{\nu_1}(z)$ in terms
of $H_{-\nu_1-1}(iz)$ and $H_{-\nu_1-1}(-iz)$, since all three functions are solutions of the same
second-order differential equation (see Ref. \onlinecite{spec-functions}). We also employ the
solution symmetry and the recurrence relation $\frac{d}{dz}H_{-\nu_1}(iz)=-2i\nu_1
H_{-\nu_1-1}(iz)$ for Hermite functions.

The integration with the under-the-barrier function can be safely extended to $u=0$, since the
deviation of the approximate integrand from the actual function in a short region of length
\mbox{$\sim\sqrt{\kappa}$} is finite. Then the integration is performed exactly in terms of the
hypergeometric function:
\begin{align}
\label{under-barrier-integral}
&
\int\limits_0^{\pi-u_1}\!\!\!du\left(\sin \frac{u}{2}\right)^{\nu}\!
\left(\cos\frac{u}{2}\right)^{s-\nu-\frac{1}{2}}\!\!=
%\\
%\nonumber
%&
\frac{\Gamma\left(\frac{\nu+1}{2}\right)\Gamma\left(\frac{-\nu_1}{2}\right)}
{\Gamma\left(s+\frac{1}{2}\right)} +
\\
\nonumber
&
\frac{2\,_2F_1\left(1,s+\frac{1}{2},\frac{2-\nu_1}{2},\sin^2\frac{u_1}{2}\right)}
{\nu_1\left(\sin\frac{u_1}{2}\right)^{\nu_1}}
\left(\cos\frac{u_1}{2}\right)^{\nu+1}\,.
\end{align}

We further expand Eq.~(\ref{other-minimum-integral}) in $\sqrt{\kappa}/u_1 \ll 1$ and
Eq.~(\ref{under-barrier-integral}) in $u_1\ll 1$. In the sum of two pieces of the integral
(\ref{R0-conjugate-integral}), the strongest dependence on $u_1$ cancels out. For the lowest
resonance with $\nu=0$ there are no other contributions comparable with the constant in
Eq.~(\ref{under-barrier-integral}), which is not the case if $\nu>0$.

As $\nu$ increases the divergence of the under-barrier solution becomes stronger, and one should
take into account all $u_1$-dependent terms with non-positive real parts of the exponents that
appear in Eqs.~(\ref{other-minimum-integral}) and (\ref{under-barrier-integral}). For large $\nu$
this requires higher order of the perturbative calculations. One might have argued that, since the
corrections to the wave functions have the same structure as the initial approximations for them,
all relevant $u_1$-dependent terms should cancel out, and the corrections to the constant are small
in $\kappa$.

However, we can avoid exact unpleasant calculations by applying the method developed above. This is
achieved by relating $(R_{\lambda_\nu^*,0})^*$ to the integral that can be calculated using the
principal approximation for the eigenfunction specified in Eqs.~(\ref{harmonic-zero} -
\ref{nu1-0}). According to the definition (\ref{P-definition}) of $P_{j,\lambda}$ and due to the
wavefunction symmetry (\ref{conjugation-symmetry}), we have
$(R_{\lambda_\nu^*,0})^*=(P_{0,\lambda_\nu^*})^*$. The recurrence relation (\ref{P-recurrence})
yields
\begin{align}
\label{second-int-recurrence}
 & (P_{0,\lambda_\nu^*})^*=
 \frac{\Gamma\left(s+\frac{\nu+3}{2}\right)\Gamma\left(s-\frac{\nu}{2}\right)}
 {(-2)^{1+\nu/2}\Gamma\left(s+\frac{1}{2}\right)\Gamma(1+s)}
\left(P_{1+\nu/2,\lambda^*_\nu}\right)^*
.
\end{align}
Next, we consider the integral
\begin{align}
&
\left(P_{1+\nu/2,\lambda^*_\nu}\right)^*=
\\
&
\nonumber
 (-1)^{1+\nu/2}\int\limits_{-\pi}^{\pi}du\,
 e^{-\frac{\cos u}{\kappa}}(1+\cos u)^{1+\nu/2}\xi_{\lambda_\nu}(u)
\end{align}
and notice that for its calculation it is sufficient to take the under-barrier approximation
(\ref{under-barrier-positive-half}) for $\xi_{\lambda_\nu}$ on the whole half-circle. The resulting
integral can be easily calculated:
\begin{align}
\label{second-int-integral}
&
\left(P_{1+\nu/2,\lambda^*_\nu}\right)^*=
\\
\nonumber
&
A_{2,\nu} (-1)^{1+\nu/2} 2^{(\nu+2s+3)/2}
\frac{\Gamma\left(\frac{1+\nu}{2}\right)\Gamma(1+s)}{\Gamma\left((2s+\nu+3)/2\right)}
\,.
\end{align}
The integral converges, and the differences between the approximate and exact integrands in the
narrow regions of width $\sim\sqrt{\kappa}$ near $u=0$ and $u=\pi$ are smaller than $\sim
A_{2,\nu}\kappa^{\nu/2}$ and $\sim A_{2,\nu}\kappa^{s+1/2}$, respectively. The calculation of the
second integral is completed by combining Eqs. (\ref{second-int-recurrence}) and
(\ref{second-int-integral}) followed by expressing $A_{2,\nu}$ via $A_{0,\nu}$, according to
Eq.~(\ref{A2}):
\begin{align}
\label{R-conjugate-result}
(R_{\lambda_\nu^*,0})^*=
%\\
A_{0,\nu} e^{-\frac{1}{\kappa}}\kappa^{-\nu/2} 2^{1+2\nu}
\frac{\Gamma\left(\frac{\nu+1}{2}\right)\Gamma\left(\frac{2s-\nu}{2}\right)}
{\Gamma\left(s+\frac{1}{2}\right)}
\,.
\end{align}
The method is naturally also applicable to the case $\nu=0$, which has been treated earlier, using
a more straightforward approach, with much more effort involved.

We complete the calculation by inserting Eqs.~(\ref{R-result}) and (\ref{R-conjugate-result}) into
Eq.~(\ref{cal-Qn}) to find that ${\cal Q}_{\lambda_\nu,0}$ and ${\cal Q}_{\lambda^*_\nu,0}$ have
finite limits at $\kappa\to 0$, namely
\begin{align}
\nonumber
&
\lim\limits_{\kappa\to 0}{\cal Q}_{\lambda_\nu,0}\
=
\frac{\Gamma(\frac{1+\nu}{2})}{\pi(\nu/2)!}
\frac{\Gamma(\frac{1-2s+\nu}{2})}{\Gamma(1+\frac{\nu-2s}{2})}
\cot (\pi s)\,,
\\
\label{limit-cal-Q0}
&
\lim\limits_{\kappa\to 0}{\cal Q}_{\lambda^*_\nu,0}=
(\lim\limits_{\kappa\to 0}{\cal Q}_{\lambda_\nu,0})^*\,.
\end{align}
We conclude the section by demonstrating that the spectral decomposition of the linear response in
the noiseless limit $\kappa \to 0$ reproduces the asymptotic expansion of its purely deterministic
counterpart presented in Ref. \onlinecite{MalininChernyak}. To that end we show that in the
noiseless limit ${\cal Q}_{\lambda_\nu,0}$ and ${\cal Q}_{\lambda^*_\nu,0}$ coincide with the
coefficients in the expansion
\begin{align}
 \label{An-expansion}
 & A_n(t;s) = \!\! \sum\limits_{\nu=0,2,\ldots} \!
 \left(Q_{\nu,n}e^{st}+\tilde Q_{\nu,n}e^{-st}\right)
 e^{-(\nu+1/2) t}
\end{align}
at $n=0$, where $A_n(t;s)$ is defined as a matrix element
of the Liouvillian deterministic evolution:
\begin{align}
\label{An-definition}
&
A_n(\zeta t;s)=\int d{\bm x}\,\psi_n^*({\bm x};s) e^{-\hat L t}\psi_0({\bm x};s)\,.
\end{align}
The integral can be calculated using the representation on the circle by solving the $1D$ Liouville
equation $\partial_t e^{-\sigma_1 t}\Psi_0(u)=-\sigma_1 e^{-\sigma_1 t}\Psi_0(u)$ [see Refs.
\onlinecite{MalininChernyak,BalazsVoros}]:
\begin{align}
\label{An-result}
 & A_n(t;s)=\frac{2\left(e^{-2t}-1\right)^n \Gamma(n+\frac{1}{2}-s)}
 {\sqrt{\pi}\Gamma(\frac{1}{2}-s)}e^{-t/2}\times
 \\
 & \nonumber {\rm Re} \Bigl( \frac{\Gamma(s)e^{st}}{\Gamma(n+\frac{1}{2}+s)}
 \phantom{|}_2F_1\Bigl(n+\frac{1}{2}-s,n+\frac{1}{2},1-s,e^{-2t}\Bigr) \Bigr).
\end{align}
The expression is substantially simplified in the relevant for the linear response case $n=0$.
Expanding the Gauss hypergeometric function $\phantom{|}_2F_1$ in $A_0(t;s)$ into the
hypergeometric series of $e^{-2t}$ to obtain $Q_{\nu,0}$ and $\tilde Q_{\nu,0}$, we see directly
that they are indeed reproduced by the noiseless limit of the coefficients ${\cal
Q}_{\lambda_\nu,0}$ and ${\cal Q}_{\lambda^*_\nu,0}$ given by Eq.~(\ref{limit-cal-Q0}), and in the
limit $\kappa\to 0$, i.e.
\begin{align}
\label{Qn-result}
 {\cal Q}_{\lambda_\nu,0}=Q_{\nu,0}
 \qquad \textrm{and} \qquad
 {\cal Q}_{\lambda^*_\nu,0}=\tilde Q_{\nu,0}\,.
\end{align}

\section{Spectral decomposition for the second-order response: Noise regularization}
\label{section:second-order}

Spectral decomposition of nonlinear response functions in the case of finite noise is conceptually
straightforward. The general expression for $S^{(2)}(t_{1},t_{2})$, decomposed in the eigenmodes of
the Fokker-Planck operator $\hat{{\cal L}}(\kappa)$ is given by Eq.~(\ref{second-order-general2}).
In this section we demonstrate that in the noiseless limit $\kappa \to 0$ the expansion
coefficients converge to the coefficients of the long-time asymptotic expansion of the purely
deterministic $\kappa=0$ response that have been derived in our previous work
\cite{MalininChernyak}. For the sake of simplicity we focus on the $N_f=1$ case when only one
irreducible representation contributes to the dipole, i.e. $f=\psi_0(\bm x;s)$. The expression
(\ref{second-order-general2}) for the second-order response contains four matrix elements
calculated in the previous section. The fifth matrix element in the middle angular brackets
requires a careful treatment since $f_-$ includes the operator
$\sigma_1\psi_0\partial/\partial\zeta$ that acts on all momentum-dependent functions to the right.
We first perform integration over the reduced phase space, represented by the middle angular
brackets, which results in a $\zeta$-dependent expression that also includes derivatives.
Integrating over $\zeta$ by parts, and employing the symmetries, we obtain the second-order
response function in the form of the following spectral decomposition:
\begin{align}
\nonumber
\label{second-order2}
&
S_{2}(t_1,t_2)
=\sum\limits_{\lambda\mu}
\int d\zeta\,\zeta\frac{\partial\rho_0}{\partial \zeta}
e^{-\zeta\lambda t_1-\zeta\mu t_2}
\lambda
\sum\limits_{n=0}^{\infty}
(-1)^n
\times
\\
&
\nonumber
(a_n-a_{n+1})
\Biggl(
\left(n+s+\frac{1}{2}\right)
\biggl[\zeta\mu t_2+n\biggr]
{\cal Q}^*_{\mu^*,n}{\cal Q}_{\lambda,n+1}
\\
&
%\nonumber
-\left(n-s+\frac{1}{2}\right)
\biggl[\zeta\mu t_2+n-1\biggr]
{\cal Q}^*_{\mu^*,n+1}{\cal Q}_{\lambda,n}
\Biggr)
\,,
\end{align}
where ${\cal Q}_{\lambda,n}$ is given by Eq.~(\ref{cal-Qn}). The symmetric $a_n=a_{-n}$
coefficients
\begin{eqnarray}
\label{an}
a_n=
\int_{M^{3}}d{\bm x}\,\psi_{n}^{*}({\bm x};s)
\psi_{0}({\bm x};s)
\psi_{n}({\bm x};s)
\end{eqnarray}
are purely geometrical factors that do not depend on the dynamics and are \cite{MalininChernyak}.
They are all related to $a_0$ with the help of the recurrence relations
\begin{eqnarray}
\label{recurrent}
a_{n+1}=\frac{8n^2+1-4s^2}{(2n+1)^2-4s^2}a_n-\frac{(2n-1)^2-4s^2}{(2n+1)^2-4s^2}a_{n-1}
\end{eqnarray}
and, therefore, can be expressed in terms of the Laplacian operator eigenfunctions $\psi_0(\bm
x;s)$.

The spectral decomposition \ref{second-order2} is an exact expression for any given $\kappa$. The
diffusion coefficient $\kappa$ enters the expression via the eigenvalues as well as via ${\cal
Q}_{\lambda,n}$, expressed in terms of the eigenfunctions as specified in
Eq.~(\ref{cal-An-series}). In what follows we determine the leading contributions in the noiseless
limit $\kappa\to 0$.

Compared to the linear case, noise plays a more delicate role for the spectral decomposition of the
second-order response function. It provides convergence of the series over angular harmonics in the
expression for the response function that does not appear in the linear case [compare
Eqs.~(\ref{linear-response1}) and (\ref{second-order2})].

To retrieve the asymptotic behavior of ${\cal Q}_{\lambda,n}$ for $\kappa \ll 1$ and $n \gg 1$ we
apply the method developed in Section \ref{section:linear} to derive the following recurrence
relation:
\begin{align}
\label{recurrence-previous}
 & \left(\lambda - \frac{\kappa n^2}{2}\right)R_{\lambda,n}=
 \\
 &
 \nonumber
 \frac{2n+1-2s}{4}R_{\lambda,n+1}- \frac{2n-1+2s}{4}R_{\lambda,n-1}
\,.
\end{align}
Specifically we replace $\lambda\xi_\lambda$ with $\hat{\cal H}\xi_\lambda$ in the integral
representation [Eq.~(\ref{R-definition})] for $\lambda R_{\lambda,n}$ followed by integrating by
parts. Eq.~(\ref{recurrence-previous}) results in another exact recurrent relation
\begin{align}
\label{recurrence}
{\cal Q}_{\lambda,n+1}=\frac{2n-1-2s}{2n+1+2s}{\cal Q}_{\lambda,n-1}
-\frac{4\lambda-2\kappa n^2}{2n+1+2s}{\cal Q}_{\lambda,n}
\,.
\end{align}
Since only symmetric eigenfunctions contribute to the spectral decomposition, we have
 ${\cal Q}_{\lambda,-n}={\cal Q}_{\lambda,n}$, which allows to express all quantities
 ${\cal Q}_{\lambda,n}$ via ${\cal Q}_{\lambda,0}$ specified by Eqs. (\ref{cal-Qn}).

We start with the limit $\kappa \to 0$ for fixed $n$. Although the dependence of $R_{\lambda,n}$ on
$\kappa$ is singular, the coefficients ${\cal Q}_{\lambda,n}$ have well-defined limits at
$\kappa\to 0$ that are equal to the corresponding coefficients in the expansion
(\ref{An-expansion}) of the purely deterministic matrix elements $A_n(t;s)$.

This can be established in the following way. Viewing $A_n(t;s)$ as a matrix element of the
evolution operator between zero and $n$-th harmonic, we employ the identity $\sigma_1=(\sigma_+
+\sigma_-)$ and the relations between the neighboring harmonics that follow from
Eqs.~(\ref{so(2,1)-circle}) to express $\partial_t A_n(t;s)$ in terms of $A_{n-1}(t;s)$ and
$A_{n+1}(t;s)$. Finally, the result is expanded in $e^{-2t}$ according to Eq.~(\ref{An-expansion}).
For the components oscillating as $\propto e^{st/2}$ we arrive at a relation:
\begin{align}
\label{Q-recurrence}
-\lambda_\nu Q_{\nu,n}= \frac{2n+1+2s}{4}Q_{\nu,n+1}-
\frac{2n-1-2s}{4}Q_{\nu,n-1} \,,
\end{align}
which that can be viewed as the noiseless limit of the recurrence relation (\ref{recurrence}).
Combined with the already established equivalence $\lim_{\kappa\to 0}{\cal
Q}_{\lambda_\nu,0}=Q_{\nu,0}$ for the zero term, this implies the equivalence $\lim_{\kappa\to
0}{\cal Q}_{\lambda_\nu,n}=Q_{\nu,n}$ for any given $n$.

So far the equivalence has been established for the RP resonances with $\lambda_\nu=\nu-s+1/2$. The
coefficients for the other set of RP resonances can be easily found by employing the symmetry of
the recurrence relation (\ref{recurrence}) combined with
 ${\cal Q}_{\lambda^*,0}=({\cal Q}_{\lambda,0})^*$, which results in:
\begin{align}
{\cal Q}_{\lambda^*,n}=\frac{\Gamma(n+\frac{1}{2}-s)\Gamma(\frac{1}{2}+s)}
{\Gamma(n+\frac{1}{2}+s)\Gamma(\frac{1}{2}-s)}
({\cal Q}_{\lambda,n})^*
\,.
\end{align}
in the limit $\kappa\to 0$. This establishes the equivalence $\lim_{\kappa\to 0}{\cal
Q}_{\lambda^*_\nu,n}=\tilde Q_{\nu,n}$ for the other set of resonances.

We are now in a position to demonstrate that the series in angular harmonics $n$ for any
coefficient in the spectral decomposition converges for small $0 < \kappa \ll 1$ and in the
noiseless $\kappa \to 0$ limit reproduces the corresponding coefficient in the long-time asymptotic
series for the purely deterministic response function $S^{(2)}(t_{1},t_{2})$.

Explicit expressions for the coefficients $Q_{\nu,n}$ and $\tilde Q_{\nu,n}$ that enter the
expansion (\ref{An-expansion}) for $A_n(t;s)$ become increasingly lengthy as $\nu$ grows. The
coefficients can be obtained by expanding the deterministic evolution matrix element in
Eq.~(\ref{An-result}) in powers of $e^{-2t}$. The simplest expressions can be found for the lowest
modes with $\nu=0$ whose energies are $\lambda=\pm s+1/2$. For example, for the mode with
$\lambda=\lambda_0^*=s+1/2$, the recurrence relation (\ref{Q-recurrence}) implies $\tilde
Q_{0,n}=(-1)^n \tilde Q_{0,n}$, where $\tilde Q_{0,n}=Q^*_{0,n}$ and $Q_{0,n}$ is specified in the
r.h.s. of Eq.~(\ref{limit-cal-Q0}) taken at $\nu=0$.

To analyze the limit $n\to\infty$ for fixed finite $\kappa \ll 1$ we view $R_{\lambda,n}$ as the
Fourier coefficients of the smooth function $\Phi_\lambda(u)$. Consequently, ${\cal Q}_{\lambda,n}$
decay faster than any power of $n$ for $n\to\infty$, and indeed we find from the recurrence
relations the intermediate asymptotic of ${\cal Q}_{\lambda,n}$ at $1 \ll n\ll \kappa^{-1}$:
\begin{align}
\label{cal-Q-asympt}
{\cal Q}_{\lambda,n}\propto (-1)^n n^{\lambda-s-\frac{1}{2}}e^{-\kappa n^2/4}
\,.
\end{align}
For larger $n$ the recurrence relation implies  ${\cal Q}_{\lambda,n}/{\cal
Q}_{\lambda,n-1}=1/(\kappa n)$, which also leads to the decay faster than any power law as $n$
increases. Finally we compare Eq. (\ref{cal-Q-asympt}) with the asymptotic $n\gg 1$ form of the
deterministic coefficients $Q_{\nu,n}$ to derive
\begin{align}
\label{Q-regularized}
{\cal Q}_{\lambda_\nu,n}=Q_{\nu,n}e^{-\kappa n^2/4}
\end{align}
which is valid in the range $1 \ll n\ll \kappa^{-1}$ for small $\kappa$. Eq.~(\ref{Q-regularized})
means that noise provides a homogeneous cut-off for all RP resonances and the series over angular
harmonics for the spectral decomposition coefficients converges at $n\sim \kappa^{-1/2}\ll
\kappa^{-1}$ where Eq.~(\ref{Q-regularized}) holds.

It remains to be demonstrated that the resulting spectral decomposition in the $\kappa \to 0$ limit
of vanishing noise coincides with the asymptotic long-time expansion of $S^{(2)}(t_1,t_2)$ in the
deterministic case. The latter expansion has been derived in Ref. \onlinecite{MalininChernyak}.

First of all, we observe that the eigenvalues of the Perron-Frobenius operator converge in the
$\kappa\to 0$ to the factors that appear in the expansion of $A_n(t;s)$. The same factors that
appear in the spectral decomposition of linear response have been interpreted in Ref.
\cite{RobertsMuz} as RP resonances. While the linear response function may be represented in the
form of a converging series, the expansion of $S^{(2)}$ appears to be more involved. If $\kappa=0$,
for a pair of resonances, the series (\ref{second-order2}) over $n$ diverges, as clearly seen from
the power-law growth in Eq. (\ref{cal-Q-asympt}).

In the deterministic case, the time dependence of the second-order response is determined by the
converging series which contains the matrix elements $A_n(t;s)$. The expansion should be formally
performed after the series summation. The long-time expansion of $S^{(2)}(t_1,t_2)$ is an
asymptotic, rather than a converging expansion in $e^{-2t_1}$ and $e^{-2t_2}$. We have developed a
method, equivalent to regrouping, which allowed to approximate the infinite sum by a sum of a
finite number of terms. Due to the alternating character of the series, any smooth cut-off
effective for larger term numbers did not influence the result.

The noise $\kappa>0$ actually introduces a smooth cutoff in the sum over $n$ in
Eq.~(\ref{second-order2}) for a given pair of resonances $\lambda,\,\mu$. The suppressive
exponential factor is present for arbitrarily small positive $\kappa$. The series over angular
harmonics is almost alternating and converging. Therefore, the sum of the series does not depend on
the value of $\kappa$ as long as it is small. It is intuitively clear that another form of
regularizing noise would lead to the same results.

For the second-order response function we have demonstrated the equivalence of the spectral
decomposition in the limit of vanishing noise $\kappa\to 0$ with the asymptotic expansion in the
case of deterministic dynamics. In short, this follows from the equality $\lim_{\kappa\to 0}{\cal
A}_n(t;s)=A_n(t;s)$ for the evolution operator matrix elements and the property that the asymptotic
expansion of $S^{(2)}(t_1,t_2)$ originates from the expansion of $A_n(t;s)$ and the subsequent
proper summation of the apparently diverging series.

\section{Discussion}
\label{sec:discussion}

In the present manuscript we have studied linear and second-order nonlinear response of a
stochastic system obtained by adding Langevin noise to a deterministic system whose dynamics is
strongly chaotic. The deterministic system is represented by a free particle moving on a $2D$
compact Riemann surface $M^{2}$ with constant negative curvature (geodesic flow). We chose the
random Langevin force to be orthogonal to the particle momentum, so that the noise does not change
the energy, and random walk occurs on the energy shell represented by the
$3D$ reduced phase space $M^{3}$.
The stochastic dynamics has been analyzed in terms of the Fokker-Planck operator
$\hat{{\cal L}}(\kappa)=-(\kappa/2)\hat{\sigma}_{z}^{2}+\hat{L}$, where the second term stands for
the deterministic component (advection), whereas the first term describes the Langevin noise in the
form of diffusion in the momentum space with $\kappa$ being the diffusion coefficient.

The $\kappa=0$ case that corresponds to purely deterministic dynamics has been studied in our
earlier work \cite{MalininChernyak}, where we have employed strong dynamical symmetry (DS) in the
system to find the analytical solution to the problem. The Langevin noise added to the system has
been chosen in a way that it does not break down the DS. Similar to the deterministic case, where
the space of reduced phase space distributions has been decomposed into simpler components
[irreducible representations of the DS group $SO(2,1)$] invariant with respect to the Liouville
operator $\hat{L}$, the same components form invariant subspaces for our Fokker-Planck
operator $\hat{{\cal L}}(\kappa)$.

Here we have considered noise as a regularization to construct spectral decompositions
of the Perron-Frobenius operator and focused on the noiseless $\kappa \to 0$ limit of the
stochastic dynamics.
We have employed the DS to analyze the eigenvalue problem for the
Fokker-Planck operator $\hat{{\cal L}}$ whose eigenfunctions are nice smooth functions for finite
values of $\kappa$. In each irreducible component the distributions can be represented by functions
on the circle, whereas the Fokker-Planck operator is represented by a sum of the first-order Liouville
operator and a $1D$ diffusion operator $-(\kappa/2)\partial_{u}^{2}$. The deterministic dynamics
has the stable $u=\pi$ and unstable $u=0$ fixed points that represent the dynamical processes along
the stable and unstable directions, respectively. Mapping the original problem onto much simpler
$1D$ stochastic dynamics on the circle allowed for an explicit analysis of the eigenvalue problem.
Being interested in the noiseless limit we focused on small finite values $\kappa \ll 1$ and found
the relevant eigenfunctions of $\hat{{\cal L}}(\kappa)$ analytically using the WKB method where the
diffusion coefficient $\kappa$ plays the role of the square $\hbar^{2}$ of the Planck constant.

The fluctuation-dissipation theorem relates the linear response function to the two-point
correlation function calculated earlier \cite{RobertsMuz}, the latter being interpreted as an
expansion in RP resonances using the language of rigged Hilbert spaces. The long-time asymptotic
expression for the second-order response functions, has the form of a spectral decomposition over the
same set of resonances \cite{MalininChernyak}. Interpretation of the asymptotic expansion in the
nonlinear case is more involved due to the following reasons. The eigenmodes that correspond to the
RP resonances are represented by generalized, rather than smooth functions. In the linear case the
initial smooth distribution should be decomposed in the RP modes. The signal is computed by
convoluting the RP modes with the smooth dipole function. Both operations are well-defined for
generalized functions. In the nonlinear case the second interaction with the driving field
involves applying a differential operator to a generalized function followed by projecting it onto
another generalized function.
The legitimacy of the latter operation is less obvious, and has been related
to the cancellation of apparently dangerous terms \cite{MalininChernyak}.

Interpretation of nonlinear response in terms of RP resonances has been addressed in this
manuscript by considering the noiseless limit of the Langevin dynamics. In the nonzero noise
case $\kappa > 0$, the spectral decomposition of the response functions of any order is conceptually
more or less straightforward. We have demonstrated explicitly for geodesic flows that in the
$\kappa \to 0$ limit the relevant eigenvalues of $\hat{{\cal L}}(\kappa)$ converge to the RP
resonances, whereas the coefficients in the spectral decompositions of the response functions
converge to the coefficients of the asymptotic series in the purely deterministic $\kappa=0$
expressions derived in Ref. \onlinecite{MalininChernyak}.

Summarizing, RP resonances can be interpreted as the noiseless limit for the eigenvalues of
the Fokker-Planck operator
$\hat{{\cal L}}(\kappa)$, and the spectral decompositions in this limit reproduce
the long-time asymptotic series for the response functions. Note that the dynamical
$\zeta$-function can be also reproduced as the limit $\zeta(z)=\lim_{\kappa\to
0}Z^{-1}\det(z-\hat{{\cal L}}(\kappa))$. In our model the RP decomposition for the linear response
is represented by a converging series, whereas the nonlinear response is given by a non-converging
asymptotic series. The converging character of the spectral decomposition is lost when the limit
$\kappa \to 0$ is applied. Computation of the expansion coefficients in the nonlinear case requires
a delicate summation of almost sign-alternating series whose convergence is ensured thanks to the
noise.

Applications of our results are not limited to interpretations of purely deterministic quantities.
Irreversibility that shows itself in the decaying correlations appears
only when the deterministic chaotic dynamics is regularized by some kind of coarse graining. Full
``physical'' mixing always requires some diffusion mechanism. The difference between stable and
chaotic deterministic dynamics is that the diffusion-induced ``physical'' mixing is much more
efficient in a chaotic system. We consider a small fraction of the phase space that represent the
initial conditions. As the ball of initial conditions is stretched and folded back, the shape becomes
elongated along unstable directions and contracted along stable ones, while the phase space volume
remains constant.
The diffusion time scales as a square of the blurring size. Therefore, a purely diffusive
relaxation in the stable system occurs on the time scale of $\tau_{\rm reg}\sim l^2/\kappa$. In a
chaotic system the scale of the density inhomogeneity decreases with time exponentially.
Considering for simplicity a $3D$ reduced phase space of a chaotic Hamiltonian system, a ball of
size $a$ of initial conditions becomes a fettuccine-like shape. The fettucine length grows as
$ae^{\lambda t}$. The size of the system $l$ induces the folding of the unstable manifold.
We estimate the number of folds as $\sim ae^{\lambda t}/l$. Then the
characteristic ``physical'' mixing time
 $\tau_{c}\sim \lambda^{-1}\ln\left(l^3 a^{-2}(\lambda/\kappa)^{1/2}\right)$
in a chaotic system with weak diffusion is very short compared
to the stable case. Note that the ``physical'' mixing time $\tau_{c}$ can be measured in photon
echo experiments, where it represents the characteristic time scale of the photon echo decay as a
function of the delay between the exciting and the dynamics-reversing pulses.

%\acknowledgments We acknowledge the
%support through the start-up funds from WSU.

\appendix
\section{Eigenmodes of the Fokker-Planck operator}
\label{appendix:Eigenmodes}

In this section we calculate the relevant eigenmodes of the Fokker-Planck operator in a given
representation of $SO(2,1)$ labeled by $s$. They are given by symmetric solutions of the
Schr\"odinger equation on a circle with the effective Hamiltonian (\ref{Hamiltonian-on-circle}):
\begin{eqnarray}
\label{SchrEq} -\frac{\kappa}{2}\frac{d^2}{du^2}\xi_\lambda(u) +\left(\frac{\sin^2 u}{2\kappa}-s\cos
u\right) \xi_\lambda(u)= \lambda\xi_\lambda(u)
\end{eqnarray}
The Hamiltonian is self-adjoint with respect to a natural symmetric scalar product.

For our purposes we focus on the weak limit $\kappa\ll 1$. The low lying states with $|\lambda|\ll
\kappa^{-1}$ are concentrated near the the potential minima, in the vicinity of $u=0$ and $u=\pi$.
The principal approximation for the solutions of Eq.~(\ref{SchrEq}) in the regions $|u|\ll 1$ or
$|u-\pi|\ll 1$ can be found by retaining up to second order terms in the expansion of the
potential.

For $|u|\ll 1$ we arrive at the approximate equation:
\begin{eqnarray}
\label{SchrEq-near-zero-app} -\frac{\kappa}{2}\frac{d^2}{du^2}\xi_\lambda(u) +\frac{u^2}{2\kappa}
\xi_\lambda(u)= \left(\lambda+s\right)\xi_\lambda(u)\,.
\end{eqnarray}
that reproduces the well-known Schr\"odinger equation for a linear harmonic oscillator. One can
easily identify the characteristic scales $\sqrt{\kappa/\kappa}=1$ and
$(\kappa\kappa)^{1/4}=\sqrt{\kappa}$ of the energy and length, respectively. A general solution of
Eq.~(\ref{SchrEq-near-zero-app}) can be represented in terms of the Hermite functions as
\begin{align}
\label{harmonic-zero-app}
\xi(u)=A_0e^{-u^2/2\kappa}H_\nu\left(\frac{u}{\sqrt{\kappa}}\right)
+B_0e^{u^2/2\kappa}H_{-\nu-1}\left(i\frac{u}{\sqrt{\kappa}}\right)\,,
\end{align}
with $A_0$ and $B_0$ being complex constants. The parameter $\nu$ is related to the energy
$\lambda$ by
\begin{eqnarray}
 \label{nu}
 \nu=\lambda-\frac{1}{2}+s\,.
\end{eqnarray}
As opposed to the case of a harmonic oscillator, the general solution (\ref{harmonic-zero-app})
contains both decaying and growing waves. Their relative amplitude can be determined only by
solving the equation on the whole circle. The functions $H_\nu(z)$ and $e^{z^2}H_{-\nu-1}(iz)$ that
are linearly independent for any parameter $\nu$ can be simply related to confluent hyperbolic
functions and parabolic cylinder functions, the latter being the solutions of the original
Schr\"odinger equation with the harmonic potential. We prefer to deal with the Hermite functions
rather than with other special functions since the former reproduce Hermite polynomials in the case
of integer order.

The general solution of Eq.~(\ref{SchrEq-near-zero-app}) near $u=\pi$ has the form similar to
Eq.~(\ref{harmonic-zero-app}):
\begin{eqnarray}
\label{harmonic-pi-app}
\xi(u)
&=&
A_1e^{-(u-\pi)^2/2\kappa}H_{\nu_1}\left(\frac{u-\pi}{\sqrt{\kappa}}\right)
\\
\nonumber
&
+&
B_1e^{(u-\pi)^2/2\kappa}H_{-\nu_1-1}\left(i\frac{u-\pi}{\sqrt{\kappa}}\right)\,,
\end{eqnarray}
where
\begin{eqnarray}
\label{nu1}
\nu_1=\nu-2s=\lambda-\frac{1}{2}-s\,,
\end{eqnarray}
and $A_1$ and $B_1$ are two complex constants.

We begin with the construction of a symmetric solution near $u=\pi$. The requirement of the
solution to be invariant with respect to $(u-\pi)\leftrightarrow (\pi-u)$ combined with the
relations between Hermite functions leads to an identity
\begin{eqnarray}
\label{BoverA-app}
\frac{B_1}{A_1}=-i\frac{2^{\nu_1}\sqrt{\pi}}{\Gamma(-\nu_1)\cos\frac{\pi\nu_1}{2}}\,,
\end{eqnarray}
and after some transformations of the Hermite functions we arrive at an explicit form of a
symmetric solution for (\ref{harmonic-pi-app}):
\begin{eqnarray}
\label{harmonic-pi1}
\xi(u)
=
&&
A_1
\frac{2^{\nu_1}\Gamma(\nu_1+1)}{\sqrt{\pi}}
%e^{-(u-\pi)^2/2\kappa}
e^{\frac{(u-\pi)^2}{2\kappa}}
e^{-i\pi\nu_1/2}
\times
\\
\nonumber
&&
\left(
H_{-\nu_1-1}\left(i\frac{u-\pi}{\sqrt{\kappa}}\right)+
H_{-\nu_1-1}\left(-i\frac{u-\pi}{\sqrt{\kappa}}\right)
\right)
\,.
\end{eqnarray}
Therefore, in the region $|u-\pi|\ll 1$ the symmetric solution is determined by two unknown
parameters $\nu\equiv\nu_1+2s$ and $A_1$; both can assume complex values.

The harmonic approximation for the potential can be used if $|u-\pi|\ll 1$, hence it is the region
where the solution given by Eqs. (\ref{harmonic-pi-app}, \ref{BoverA-app}) is valid.

The general WKB solution of Eq.~(\ref{SchrEq}) under the barrier and around $u=\pi/2$ is
represented by a superposition of two waves:
\begin{align}
\label{under-barriers}
\xi(u)=\frac{A_2}{\sqrt{p(u)}}\exp(-S(u))+\frac{B_2}{\sqrt{p(u)}}\exp(S(u))\,,
\end{align}
where $p(u)$ and $S(u)$ are defined by
\begin{eqnarray}
&&
p(u)=\sqrt{\sin^2u-2\kappa s \cos u - 2\kappa\lambda} \,,
\\
\label{S2S3}
&&
S(u)=\frac{1}{\kappa}\int\limits_{\pi/2}^u dw\,p(w)
\,.
\end{eqnarray}
The quasiclassical expansion over $\kappa$ for the wavefunction phase can be employed when the
wavelength $\kappa/p(u)$ does not change too fast:
\begin{eqnarray}
\left|\frac{d}{du}\left(\frac{\kappa}{p(u)} \right)\right| \ll 1\,.
\end{eqnarray}
This holds when we are not too close to the classical turning points where $p(u)$ turns to zero.
Small value of $\kappa$ ensures that the classical turning points lie close to $u=0$ or $u=\pi$.
The WKB solutions under the barriers are valid at least for $|u|\gg\sqrt{2\kappa(\lambda+s)}$ and
$|u-\pi|\gg\sqrt{2\kappa(\lambda-s)}$. The right-hand sides of these inequalities contain
expressions for the classical turning points.
If $|\lambda|,|s|\sim 1$, the validity of the quasiclassical approximation
(\ref{under-barriers}) conditions is limited to the regions
\begin{eqnarray}
\label{under-barrier-conditions}
|u|,|u-\pi|\gg\sqrt{\kappa}\,.
\end{eqnarray}

We can invoke a further approximation in the expressions for the solutions under the barrier. For
$\kappa\ll 1$ and $|\sin u| \gg \sqrt{\kappa}$ (i.e. in the regions determined by the same
inequalities (\ref{under-barrier-conditions})), we arrive at a simplified expression for $S(u)$,
where only those terms that are small compared to one have been neglected:
\begin{align}
\label{S2-approximation}
S(u)=
&
\frac{1}{\kappa}\int\limits_{\pi/2}^u dw\,
\left(\sin w-\kappa s\frac{\cos u}{\sin u} -\frac{\kappa\lambda}{\sin u}\right)
=
\\
&
-\frac{\cos u}{\kappa}-s\ln\sin u-\lambda\ln\tan\frac{u}{2}
\,.
\end{align}
In approximating the pre-exponential factor of Eq.~(\ref{under-barriers}) it is sufficient to set
$p(u)=\sin u$.

\begin{figure}[ht]
\centerline{
\includegraphics[width=3.0in]{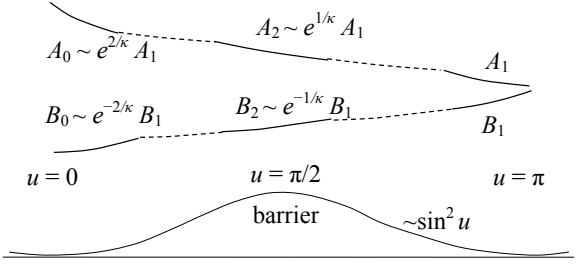}
}
\caption{ Schematic picture of the amplitudes that result from the tunneling through the potential
barrier. \label{eigenfunction} }
\end{figure}

We can see that in all intervals the wave function consists of two components that exponentially
grow in the opposite directions under the barrier. A crude estimate for the magnitudes of the two
components is presented in Fig. \ref{eigenfunction}. Only the dominant part
$\sin^2u/(2\kappa)$ of the potential contributes to the exponential dependence of the barrier
transparency on $\kappa$.

We further notice that for $u$ close to $\pi$ (actually in the interval $\kappa^{1/2}\ll(\pi-u)\ll
\kappa^{1/4}\ll 1$) we can use the expansion
\begin{eqnarray}
\nonumber S(u)=\frac{1}{\kappa}-\frac{(u-\pi)^2}{2\kappa}+(\lambda-s)\ln (\pi-u)-\lambda\ln 2 \,,
\end{eqnarray}
where the neglected terms are small compared to one.

This is the interval where the WKB solution under the barrier and the solution in the harmonic well
both represent a good approximation and can be matched. We use the asymptotic expressions for the
Hermite functions valid for $|u-\pi|\gg \sqrt{\kappa}$. In fact, taking into account Eqs.
(\ref{nu}, \ref{nu1}), we can see that the term with $A_2$ in Eq.~(\ref{under-barriers}) and the
solution given by Eq.~(\ref{harmonic-pi}) contain the same quadratically increasing argument in the
exponential, and the same power-like pre-exponential dependence:
\begin{align}
\label{compare-asympt}
%(\pi-u)^{-\nu_1-1}
%
&
A_2 (\pi-u)^{-\lambda+s-\frac{1}{2}}
2^{\lambda}\exp\left[-\frac{1}{\kappa}+\frac{(u-\pi)^2}{2\kappa}\right]=
\\
\nonumber
&
B_1 \left(\frac{2(\pi-u)}{\sqrt{\kappa}}\right)^{-\nu_1-1}
\exp\left[-\frac{i\pi(\nu_1+1)}{2}+\frac{(u-\pi)^2}{2\kappa}\right]
\,.
\end{align}
In Eq.~(\ref{compare-asympt}) we have also used the symmetry $\xi(u)=\xi(2\pi-u)$, which implies
that for positive $(u-\pi)\gg \sqrt{\kappa}$ the asymptotic form of $\xi(u)$ (\ref{harmonic-pi})
includes the only term that is proportional to $B_1$.

Therefore, the matching near $u=\pi$ yields
\begin{align}
\label{B1-app}
B_1=A_2 e^{-\frac{1}{\kappa}} 2^{\nu_1+\lambda+1} \kappa^{-\frac{\nu_1+1}{2}} e^{\frac{i\pi(\nu_1+1)}{2}}
\,.
\end{align}
The value of $A_1$ is then found from Eq.~(\ref{BoverA-app}). The component of $\xi(u)$ in
Eq.~(\ref{harmonic-pi-app}) with $B_2$ in front of it decreases rapidly $\propto\exp(-\cos
u/\kappa)$ under the barrier as $u$ deviates from $\pi$. Its magnitude can be estimated as
\begin{align}
B_2\propto A_1 e^{-\frac{1}{\kappa}}
\,.
\end{align}

The procedure of solving the Schr\"odinger equation near $u=0$ is completely similar: In the region
$\sqrt{\kappa}\ll|u|\ll 1$ both approximate solutions (\ref{harmonic-zero-app}) and
(\ref{under-barriers}) are adequate and can be matched. To that end we expand $S(u)$ in the
interval $|u|\ll \kappa^{1/4}$, neglecting the terms that are small compared to one:
\begin{eqnarray}
\nonumber S_2(u)=-\frac{1}{\kappa}+\frac{u^2}{2\kappa}-(\lambda+s)\ln u+\lambda\ln 2 \,,
\end{eqnarray}
and also employ the asymptotic forms of the Hermite functions in the overlap region. With $u$
approaching $0$ the second term in Eq.~(\ref{under-barriers}) decreases and hence matches the
second term in Eq.~(\ref{harmonic-zero-app}),
\begin{align}
B_0\propto B_2 e^{-\frac{1}{\kappa}}\propto A_1 e^{-\frac{2}{\kappa}}
\,.
\end{align}
The component with $A_2$ in front of it matches under the barrier with the component that has $A_0$
in front of it and decreases as $u$ deviates from $0$:
\begin{align}
\label{A2-app}
A_2=A_0 e^{-\frac{1}{\kappa}} 2^{\nu+\lambda} \kappa^{-\frac{\nu}{2}}
\,.
\end{align}
First we assume that the ratio $B_1/A_1$ given by Eq.~(\ref{BoverA-app}) does not contain
exponentially small term $\sim e^{-1/\kappa}$. Then  Eqs. (\ref{B1-app})-(\ref{A2-app}) imply that
the second component in Eq.~(\ref{harmonic-zero-app}) scales $B_0\propto e^{-4/\kappa} A_0$ near
$u=0$. Its presence is related to the nonresonant tunneling along the circle from the potential
minimum into itself. The argument in the suppressive exponential factor is evaluated as
\begin{align}
\frac{1}{\kappa}\int\limits_{0}^{2\pi} du\,p(u)\approx \frac{1}{\kappa}
\int\limits_{0}^{2\pi} du\,|\sin u|=
\frac{4}{\kappa}
\,.
\end{align}
The tunneling is nonresonant, since the two wells near $u=0$ and $u=\pi$ are offset by $s$ which is
not an integer number.

Performing the same procedure in the region with $\sin u < 0$ as it has been done for $\sin u >0$,
we find the symmetric solution
\begin{align}
\label{harmonic-zero1} \xi(u)=A_0e^{-u^2/2\kappa}H_\nu\left(\frac{|u|}{\sqrt{\kappa}}\right)\,,
\end{align}
valid for $|u|\ll 1$. For arbitrary $\nu$ the function may have a cusp at $u=0$. The solution is
smooth if $\nu$ is a nonnegative even integer. Then the Hermite functions $H_\nu$ reduce to Hermite
polynomials of even order. The dependence $\lambda_\nu=\nu+(1-s)/2$ on $\nu$ with $\nu=0,2,\ldots$
of the eigenstate energy for the states concentrated near $u=0$ is given by Eq.~(\ref{nu}).
Exponentially small corrections $\propto e^{-4/\kappa}$ , induced by the tunneling can be
neglected. On the other hand, the power-like in $\kappa$ corrections to the energies, play a
crucial role for $\nu>0$. The latter corrections appear because of the deviation of trigonometric
functions from their harmonic approximations. However the method developed in Section
\ref{section:linear} allows to avoid these calculations completely.

Complex conjugation of the Hamiltonian that corresponds to the change of sign of $s$, or to the
shift $u\to u+\pi$ implies the quantization condition $\nu_1=2k$ with $k=0,1,2,\ldots$ for the
other set of eigenstates. They have energies $\lambda=2k+(1+s)/2$ and are concentrated near
$u=\pi$. The eigenfunctions satisfy the relations similar to the specified above.

The eigenfunctions of the Hamiltonian (\ref{Hamiltonian-on-circle}) found above significantly
differ from those of the linear harmonic oscillator only far from the interval $|u|\lesssim
\kappa^{1/4}$. Outside the interval the exponentially decaying functions take negligibly small
values so that the normalization factor has the form
typical to the harmonic oscillator:
\begin{align}
\label{A0-app}
A_{0,\nu}=(\pi\kappa)^{-1/4}(2^\nu \nu!)^{-1/2}\,.
\end{align}
The phase of the normalized eigenfunctions of Hermitian operators can be chosen arbitrarily. For
our Hamiltonian and scalar product (\ref{scalar-product-real}) this reduces to the freedom in the
sign choice.

Finally, for illustrative purposes, we calculate the correction to the energy
$\lambda_\nu=\nu-s+1/2$ of the eigenstate that originates from the vicinity of $u=0$ ( note that
the result is not used anywhere):
\begin{align}
\label{do-not-need-it} &
\lambda_\nu^{(1)}=\int\limits_{-\pi}^{\pi}du\,\left(\frac{su^2}{2}-\frac{u^4}{6\kappa} \right)
\xi^2_{\lambda_\nu}(u)=
\\
&
\frac{\kappa}{8}(2s(1+2\nu)-(1+2\nu+2\nu^2))
\,.
\end{align}
One of the ways derive Eq.~(\ref{do-not-need-it}) is to consider the next term in the WKB expansion
under the barrier. The WKB wavefunction contains the state energy. One can simply require that the
wavefunction does not contain logarithmic terms due to their absence near $u=0$.

\end{document}